# AN IMPROVED PHYSICS BASED NUMERICAL MODEL OF TUNNEL FET USING 2D NEGF FORMALISM

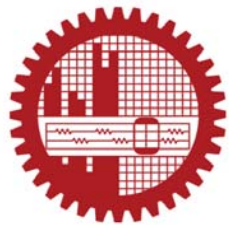

A thesis submitted to the
Department of Electrical and Electronic Engineering (EEE)
of
Bangladesh University of Engineering and Technology (BUET)

In partial fulfillment of the requirement for the degree of
Bachelor of Science in Electrical and Electronic Engineering

Submitted by
**MD. SHAMIM HUSSAIN**
**(ID: 1106032)**

Department of Electrical and Electronic Engineering (EEE)
Bangladesh University of Engineering and Technology (BUET)
February 2017

# CERTIFICATION

The thesis titled "An Improved Physics Based Numerical Model of Tunnel FET Using 2D NEGF Formalism" submitted by Md. Shamim Hussain (1006032), has been accepted satisfactory in partial fulfillment of the requirement for the degree of Bachelor of Science in Electrical and Electronic Engineering on, _____________________2017.

Supervisor:

_______________________

Dr. Md. Kawsar Alam

Associate Professor

Department of Electrical and Electronic Engineering

Bangladesh University of Engineering and Technology

Dhaka-1000, Bangladesh.

ii

# DECLARATION

I confirm that, the work presented in this thesis entitled "An Improved Physics Based Numerical Model of Tunnel FET Using 2D NEGF Formalism" is the outcome of the investigations carried out by me under the supervision of Associate Professor Dr. Md. Kawsar Alam in the Department of Electrical and Electronic Engineering, Bangladesh University of Engineering and Technology (BUET), Dhaka. I also declare that, neither this thesis nor any part thereof has been submitted or is being currently submitted anywhere else for the award of any degree or diploma.

\_\_\_\_\_\_\_\_\_\_\_\_\_\_\_\_\_\_\_\_\_\_\_\_\_\_

Md. Shamim Hussain

Student ID: 1106032

*"I was born not knowing and have had only a little time to change that here and there"*

**-Richard P. Feynman**

# Abstract

In recent years, Tunneling Field Effect Transistor (TFET) have attracted a lot of attention due to their remarkable attributes like lower than 60mV/decade subthreshold swing and higher scalability. TFETs are specially promising because of their ability to reduce the power dissipation in electronic chips like microprocessors and their suitability to be integrated in existing fabrication technology. As different designs of TFET are being investigated, proper simulation tools are essential to predict and improve their performance levels. Although semi-classical TCAD based simulation tools are predominantly used for quick evaluation of a design, these methods suffer from several limitations due to their simplistic nature, the most prominent of which are the difficulty to predict tunneling paths and the failure to properly take into account quantum mechanical effects. Although, elaborate quantum mechanical simulations like k.p and tight-binding simulations can go beyond these limitations, they are also significantly more complex and require a lot more computational resources and time. Although, 2 band models based on 1D NEGF have been proposed and investigated in some works previously, an appropriate 2D model for quantum ballistic transport is yet to be reported. In this work we have investigated a 2D model of band to band tunneling based on 2 band model and implemented it using 2D NEGF formalism. Being 2D nature, this model better addresses the variation in the directionality of the tunneling process occurring in most practical TFET device structures. It also works as a compromise between semi-classical and multiband quantum simulation of TFETs. In this work we have presented sound step by step mathematical development of the numerical model. We have also discussed how this model can be implemented in simulators and pointed out a few optimizations that can be made to reduce complexity and to save time. Finally, we have performed elaborate simulations for a practical TFET design and compared the results with commercially available TCAD simulations, to point out the limitations of the simplistic models that are frequently used, and how our model overcomes these limitations.

# Acknowledgements

I would like to express my sincere gratitude to my supervisor Dr. Md. Kawsar Alam, Associate Professor of the department of Electrical and Electronic Engineering, BUET for his continuous support, patience, motivation and immense knowledge. His guidance helped me all the time of the research and in writing of this thesis. He shared his knowledge with me in interpreting subject topics and also valued my way of thinking to synthesize those topics. His suggestions pushed me towards the direction of better thinking, his brilliant reviews refined me in working out my research problems, and his support gave me spirit to continue my work at the time of disappointment. I am extremely thankful and indebted to him for sharing his expertise and for offering his valuable guidance and encouragement.

I wish to express our sincere thanks to Dr. Quazi Deen Mohd Khosru, Head of the Department, for providing me with all the necessary facilities for the research. I take this opportunity to express gratitude to all of the Department faculty members for their help and support. Their valuable reviews and remarks helped me a lot in refining my work.

I would deeply thank parents and my elder sister for the unceasing encouragement, support and attention they provided me with. They always believe in me and inspire me to work for good. I am grateful to them for their never-ending support and inspiration.

I also place on record, my sense of gratitude to one and all, who directly or indirectly, have lent their hand in this thesis work.

# Table of Contents

# List of Figures

# List of Tables

# Chapter 1
# Introduction

In this chapter, we briefly introduce the history of integrated circuits and review the scaling of transistors in chips that has driven the electronics industry. Then we point out the limitations of MOSFET when it comes to scaling and optimizing power density on chip. We introduce Tunnel FET as a promising alternative to MOSFET which can circumvent the problems faced in scaling. We mention the various advantages of TFET, review the work that has already been done in literature on TFET and demonstrate the variety in TFET design structures. Finally, we comment on the factors that motivated us to carry out this work of thesis.

## 1.1 The Scaling Trend of Transistors

The quest for high performance computers, better communication equipment and consumer electronics like cellphones have been the driving force behind the advancement of integrated electronics. To increase performance, speed and reduce power and cost, the scaling of transistors have been the main trend in electronics, besides increasing the chip size and optimization of logic architecture. As the feature size is shrunk down, more transistors can be packed together in a chip, which in turn results in lower cost and better performance [1].

The trend of shrinking transistor size can be summarized by Moore's law [2] proposed by Gordon Moore in 1965, which predicts that the number of transistors on a dense integrated electronic chip doubles approximately every two years. The device performance doubles approximately every three years. Remarkably, for 50 years, the electronics industry has kept pace with Moore's law [3], and the feature size have been shrunk down to tens of nanometers.

Earlier in the electronics industry, or more specifically the CMOS IC industry, constant voltage scaling was the trend [1], where power the supply voltage was kept constant. In that period the performance of a chip was signified basically by its speed; power consumption was considered a lesser concern because there were not as many transistors on the chip as modern ICs. The constant voltage scaling allowed the chips to rapidly increase in speed, at the cost of higher power dissipation, but soon this became a problem. The transistors were being shrunk down, this implied they were being packed together more densely. This resulted in higher power density on the chip and was threatening to cause thermal runaway and chip failures. The power to performance ratio also became an issue at the onset of mobile devices like cellphones and laptop computers, where battery life is a major concern. So, as shown in Figure 1.1 the CMOS IC technology switched to constant field scaling, which incorporated MOSFETs operating at lower supply voltage, which kept the power density on the chip constant and allowed the scaling trend to continue.

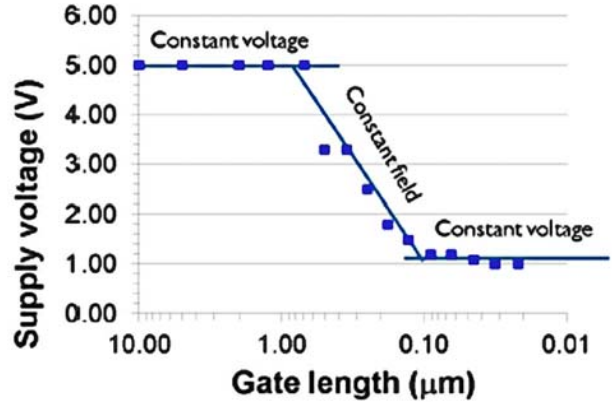

**Figure 1.1: The scaling strategy adopted for different gate lengths**

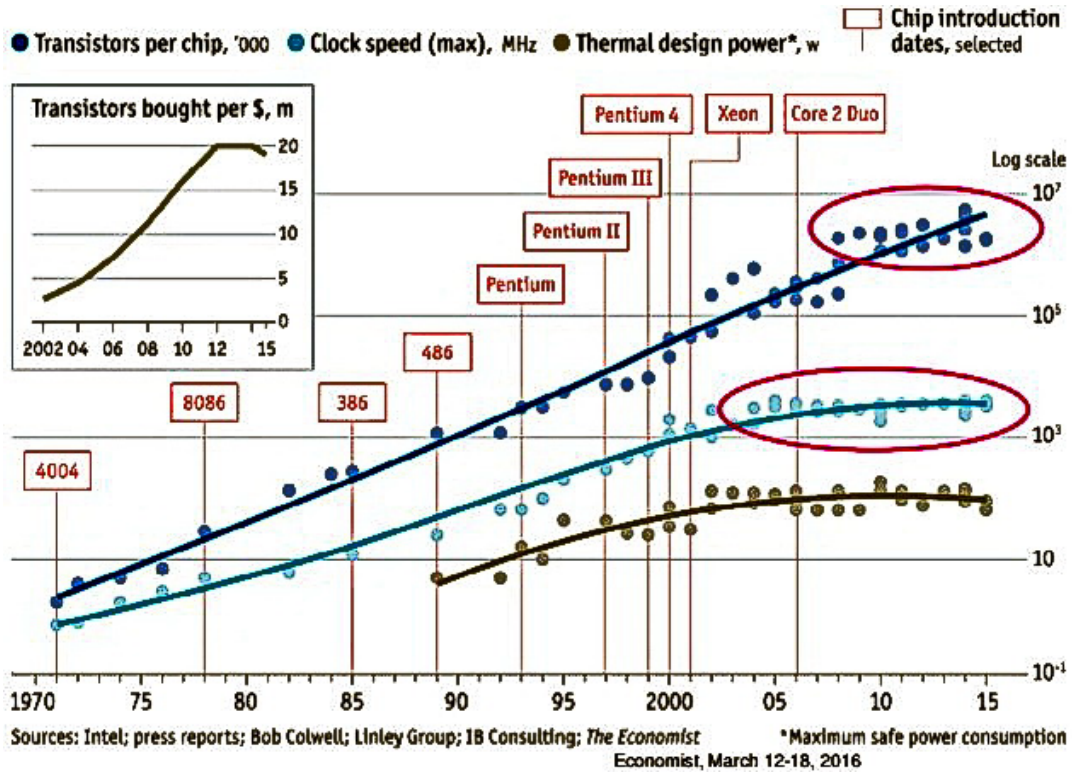

**Figure 1.2: Demonstration of deviation from Moore's law for Intel microprocessors [4]**

However the reduction of supply voltage requires that we reduce the threshold voltage of the MOSFETs as well, but it cannot be done indefinitely due to the laws of thermodynamics [5]. So, ultimately the limitations imposed by physics has forced us to adopt constant voltage scaling again and consequently the trend of scaling has slowed down since the 65 nm node. This problem is made more complicated by the appearance of quantum mechanical phenomena like the short channel effects and gate leakage current [6].

Inevitably, due to the slowdown of scaling, all the performance parameters – transistors per chip, clock speed and transistors bought per dollar have undergone a downward trend from Moore's law [4]. This is evident from the data collected for Intel processors shown is Figure 1.2. If we want to keep pace with Moore's law we must introduce new transistor technologies that will supplement or replace MOSFETs in logic circuits.

## 1.2 Limitations of MOSFET

A MOSFET uses electric field to attract thermally generated electrons (or holes) and thus form a conducting channel. In other words, electrons travel from source to drain over a barrier which is controlled by the gate. The transmission of carriers over the barrier follows the laws of thermionic emission in classical thermodynamics. When the MOSFET channel is scaled down below roughly 65 nm, quantum phenomena like source to drain tunneling become prominent which hamper the gate control.

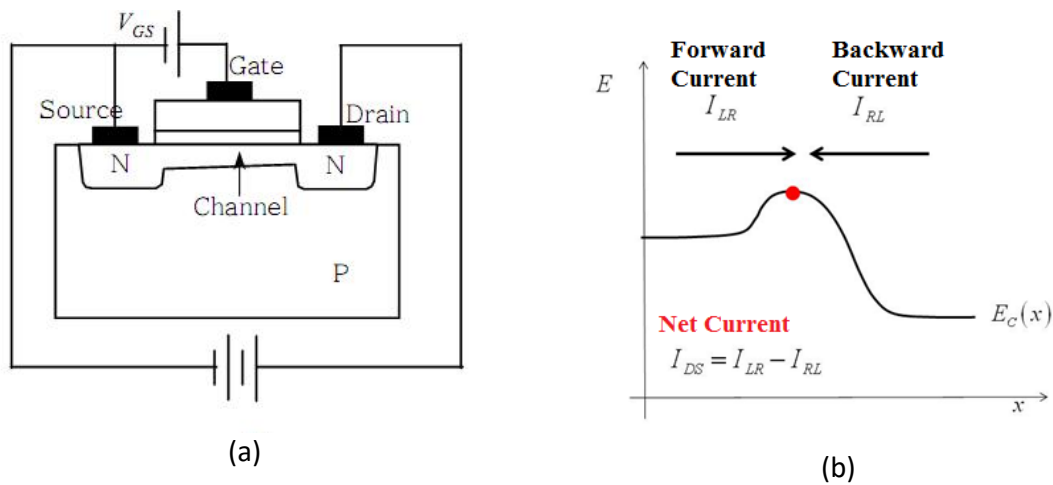

**Figure 1.3: The operating principle of MOSFET (a) creation of a conducting channel at on state and (b) thermionic emission over the channel barrier and the resulting currents**

To explain why the threshold voltage of MOSFETs cannot be reduced below a certain limit we first introduce the concept of Subthreshold Swing (SS). It signifies how steep the turn on curve is for a transistor i.e. how fast the current increases with increase in gate voltage from off state to on state.

$$SS = \frac{dV_{GS}}{d\log(I_D)} = \frac{dV_{GS}}{d\ln(I_D)} \times \ln(10) \tag{1.1}$$

The subthreshold drain current for a MOSFET is given by [7]

$$I_D(sub) \propto \left[\exp\left(\frac{qV_{GS}}{kT}\right)\right] \tag{1.2}$$

Where, $V_{GS}$ is the gate to source voltage and $kT/q$ is the thermal voltage, which is 25.9 mV at room temperature. This relation implies that the subthreshold swing of MOSFETs can never be below 60mV/decade at room temperature. This is a physical limit, which cannot be overcome even by modification of the basic MOSFET structure [5].

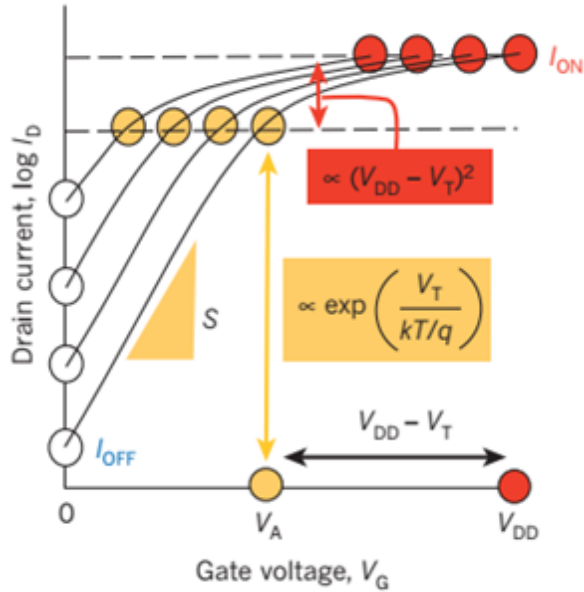

**Figure 1.4: Demonstration of increase in off leakage drain current due to reduction of threshold voltage as the SS stays constant [8]**

The dynamic power dissipation is given by [1]

$$P_{dynamic} = f C_{ox} V_{DD}^2 \tag{1.3}$$

Where, $f$ is the clock frequency, $C_{ox}$ is the oxide capacitance and $V_{DD}$ is the supply voltage. So, to reduce, dynamic power dissipation on a chip we want reduce supply voltage. But, reducing the power supply demands that we reduce the threshold voltage to obtain sufficient on current. However, the limitation on the subthreshold swing of MOSFET implies that the off current (or leakage current) will also increase with decreasing threshold voltage and subsequently lead to higher static power dissipation [1], which is given by:

$$P_{static} = I_{off} V_{dd} \tag{1.4}$$

Ultimately high static power dissipation offsets the improvement in dynamic power dissipation [8], and no further improvement in power density on a chip can be obtained by MOSFETs.

## 1.3 Beyond MOSFET: The Tunnel FET

The Tunneling Field Effect Transistor or Tunnel FET (TFET) is a three terminal device. Just like conventional transistors like MOSFET their source to gate current is controlled by the voltage applied at the gate. But TFETs use a different type of mechanism for turning on viz. Band To Band Tunneling (BTBT). Band to band tunneling is the process in which the electron tunnels from valence to conduction band through the bandgap or vice versa [9]. Band to band tunneling is a quantum mechanical phenomenon which is very different from the thermionic emission in MOSFETs. BTBT current depends mainly on tunneling distance and barrier height (potential energy); not so much on channel conductance.

TFETs are structurally similar to MOSFETs as shown in Figure 1.5(a), having source channel and drain regions, except their doping types are different (generally p-i-n). The drain current results from tunneling injection of electrons from valence band to conduction band. The tunneling distance is modulated by the applied gate voltage which changes the electric field and thus the bending of the bands at the channel region.

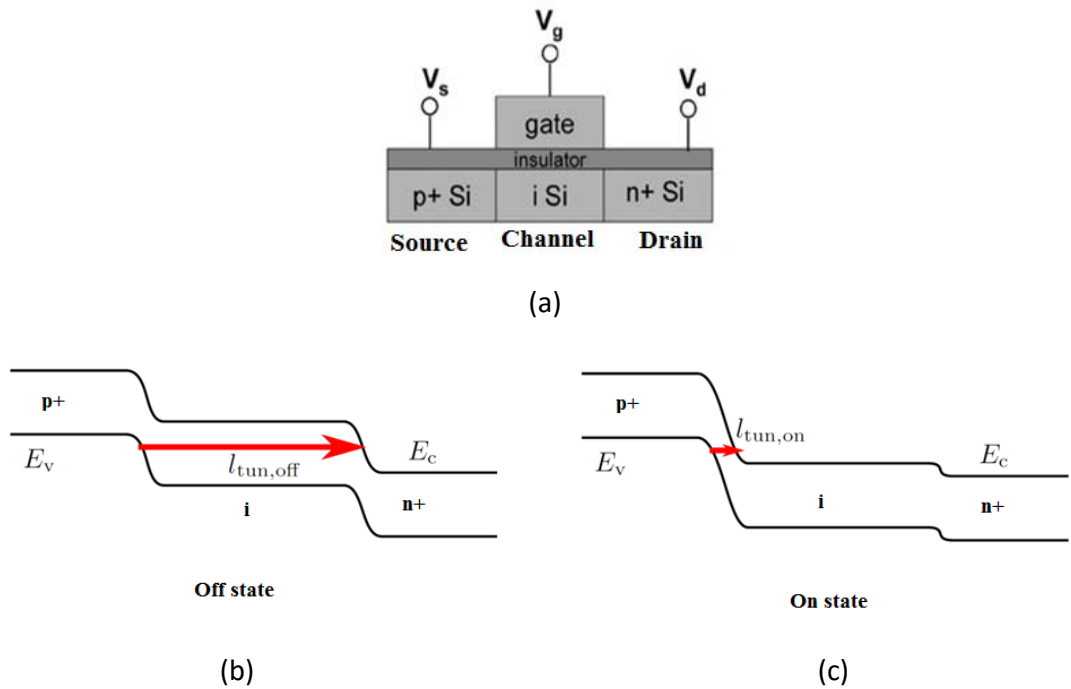

Figure 1.5: (a) Basic structure of a MOSFET (b) Long tunneling distance at off state (c) Short tunneling distance at on state

The transfer of electron from valence to conduction band by BTBT can be thought of as an electron hole pair generation process. At the off state the tunneling distance is long and tunneling current is almost zero (Figure 1.5(b)). The device acts like a reverse biased diode, so the off current is very low. At on state, electron need to tunnel a much shorter distance (Figure 1.5(c)), so tunneling current is high which appears as the drain current.

## 1.4 Comparison between TFET and MOSFET

One of the most important feature of TFET is that, in TFET electron-hole pairs generated by band to band tunneling causes the drain current, not the thermally generated carries. This process is described by the laws of quantum mechanics, rather than thermodynamics. As current depends on tunneling process, it is possible to achieve lower subthreshold swing, even lower than the thermodynamic limit of 60 mV/decade in MOSFETs which has been experimentally demonstrated [10]. So, TFETs have lower threshold voltage and can operate with lower off leakage current at lower supply voltages, contrary to MOSFETs [11]. However, for proper operation of TFETs, we need a rapid change in band profile with changing gate voltage which requires abrupt change in doping profiles , which were not possible before the introduction of modern fabrication technologies.

The on current of TFET is typically much lower than that of MOSFET because this current is produced by the tunneling process which usually cannot produce as much current as the thermionic emission process for the same applied gate voltage. So, generally TFETs outperform MOSFETs mainly at the lower power regime [8], where drain current requirement is moderate or low. This is the requirement in Mobile devices like cellphones or laptop computers.

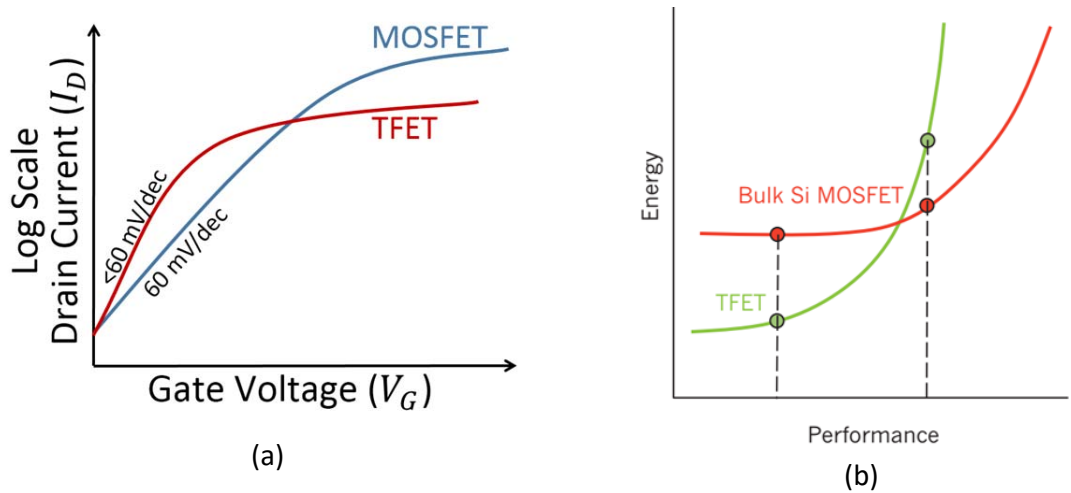

**Figure 1.6: Comparison between MOSFET and TFET [8] (a) Drain current vs Gate voltage; MOSFET has higher on current but TFET has higher SS (b) Energy requirement vs performance; MOSFET performs better at higher power levels while TFET performs better at lower power levels.**

Before continuing with our discussion on TFETs, we must clarify some other differences between TFET and MOSFET-

- Only one type of carrier i.e. electron or hole participate in MOSFET operation, whereas both electron and holes participate in TFET operation. Thus carrier transport in TFET is ambipolar in nature [8].

- In MOSFET electrons remain in the conduction band while traveling from source to drain, but in TFET electrons travel first in the valence band, then in the bandgap (as evanescent waves) and finally in the conduction band.
- The tunneling that occurs in short channel MOSFET is intra-band tunneling and it is considered detrimental to MOSFET performance [6]. Whereas in TFET electrons travel from band to band and it is a requirement for the operation of the device.
- Carrier mobility has a huge impact on MOSFET performance because it is a transferred electron device. Whereas in TFET, the structure of the bands, i.e. the bandgap and the barriers is the major concern.

## 1.5 Advantages of Tunnel FET

TFETs are especially attractive because they can be integrated in already existing semiconductor technology like the well-established Silicon fabrication technology. In the future when compound semiconductors like the III-V and other materials find their way into the electronics market, TFETs can take advantage of the new materials.

Even though TFETs are unlikely to replace MOSFETs completely, they can work alongside MOSFETs where power economy is crucial and thus reduce the overall power consumption of the chip [12]. This will be extremely useful for mobile devices to extend their battery lives and performance levels. This will also make electronic devices greener and help reduce the global power crisis.

Unlike MOSFETs, the short channel effects are not prevalent in TFETs [12]. So, the feature size of TFETs can be much smaller than MOSFETs. TFETs can also help reduce the power density on chips, which will allow us to pack more transistors together without causing thermal runaway. This will allow us to continue the trend of scaling according to Moore's law.

Going beyond conventional semiconductor technology, nanowire TFETs like CNT TFETs show great promise [13], which can achieve very low subthreshold swing even at lower doping levels. Nanowire TFETs may one day redefine the world of electronics.

## 1.6 Literature Review

Long before the concept of tunneling device was conceived, the theory of band to band tunneling was given by Clarence Melvin Zener [14] who introduced the concept of band to band tunneling in 1934 to explain the breakdown mechanism in reverse biased p+ - n+ junctions. This breakdown mechanism is now called Zener breakdown, and this type of junctions are used in Zener diodes which are widely used in voltage regulator circuits. Band to band tunneling is often called Zener tunneling.

The first investigation of a transistor containing the basic elements of the TFET was conducted by Stuetzer [15] in 1952. Stuetzer showed the ambipolar nature of the current–voltage I–V, in the field gating of a lateral Ge p–n junction. He was also able to show the dependence of the transistor characteristic on gate placement with respect to the p–n junction.

The first demonstration of a true tunneling device was the tunnel diode by Leo Esaki [16] in 1958. These diodes show negative resistance over a particular range of forward voltage which is achieved by band to band tunneling. This characteristic of the diode is used in oscillator circuits like the microwave resonators.

After this initial work by L. Esaki in heavily doped semiconductor junctions, tunneling phenomenon was shown to play a crucial role even in metal–oxide–metal (MOM) and metal–oxide–semiconductor (MOS) diodes [9]. This concept was utilized in the first proposal for a tunnel transistor by Mead in 1960 [17]. The device is sometimes called the MOMOM (metal-oxide-metal-oxide-metal) transistor. In this case the electrons are injected from the emitter into the base via tunneling.

The first calculations of the Zener tunneling rate in a uniform electric field were made by Keldysh [18] in 1958 and Kane [19] in 1960 for a direct semiconductor based on a two band model. The problem of Zener tunneling in an indirect semiconductor was treated earlier by Keldysh [20] and Kane [21], whereas Schenk [22] formulated a model for the tunneling probability in a uniform field using the Kubo formalism [23].

In 1977, Quinn et al. [24] proposed the formation of a surface-channel MOS tunnel junction by replacing the n-type source of an n-MOSFET with a highly degenerate p-type source. Their device geometry, the configuration of a lateral TFET, was intended for measurement of subband splitting and transport properties of tunneling between a bulk source and a 2-D surface channel. The first vertical TFET appears to have been proposed by Leburton et al. [25] with the aim of creating a high-speed transistor in which the gate was used to control the negative differential resistance (NDR).

S. Banerjee [26] in 1987 first reported of a novel three terminal transistor action in a silicon trench capacitor that worked on the principle of Zener tunneling. In 1992 T. Baba [27] formally proposed the concept of a surface tunnel transistor (STT) using a gated P+-I-N+ diode. Transistor action was demonstrated in GaAs based STTs with i-Al0.6Ga0.4As as the gate dielectric. In 1995 W. M. Reddick [28] experimentally demonstrated STT action in Silicon. In 2000, W. Hansch [29] proposed and fabricated a vertical silicon tunnel transistor taking advantage of MBE grown epitaxial layers to create highly abrupt tunnel junctions. This was followed by the work of K. K. Bhuwalka [30] who proposed the use of δp+ doped SiGe layer next to the source-channel tunnel junction in a vertical device architecture to improve its drive current and sub-threshold slope.

Modern interests in TFETs is due to J. Appenzeller [10] who in 2004, experimentally demonstrated sub-60mV/decade sub-threshold slope in a carbon-nanotube based transistor, clearly explaining its functionality based on band to band tunneling. Within few years, groups from UC Berkley and Stanford demonstrated that sub-60 swing can be achieved using traditional semiconductor materials i.e Silicon and Germanium. Recently, TFETs fabricated in various material systems (carbon, silicon, SiGe and group III–V materials) have emerged experimentally as the most promising candidates for switches with ultralow standby power and sub-0.5 V logic operation [31-34].

## 1.7 Tunnel FET Structures

Over the years, many designs of TFET structure have been proposed and experimented on [35]. We will briefly discuss only a few of the most prominent designs. When designing TFETs, there is a tradeoff between performance and complexity in fabrication. TFETs can be significantly more difficult to fabricate from their MOSFET counterparts due to the requirement of abrupt junctions.

A few basic structures of TFET are shown in Figure 1.7. A brief description of these structures are given below-

- ➢ TFET designs can be lateral or vertical. Lateral TFETs bear resemblance to conventional MOSFETs, whereas vertical TFETs are grown by epitaxy.
- ➢ Lateral TFETs are generally either Single Gate (SG) or Dual Gate (DG). They can be multi-gate or gate-all-around like nanowire TFETs as well.

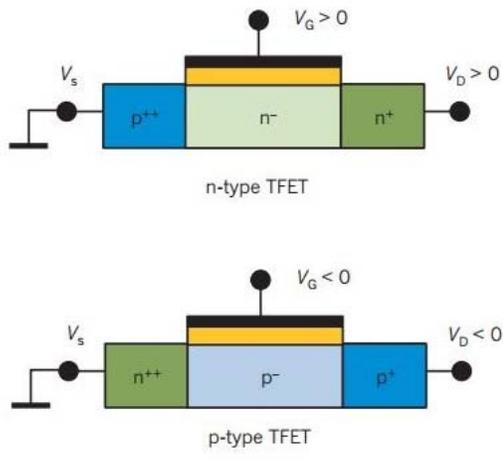
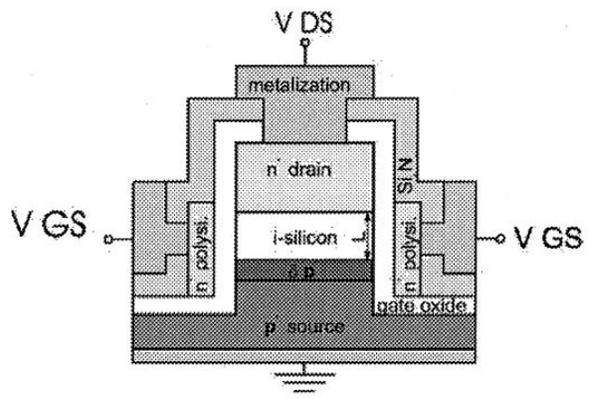

(a)                                           (b)

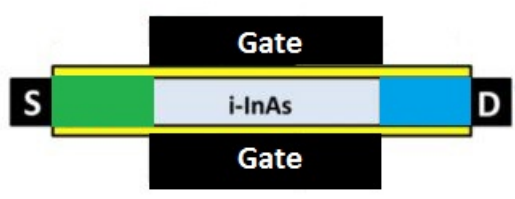
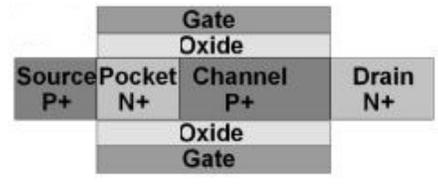

(c)                                           (d)

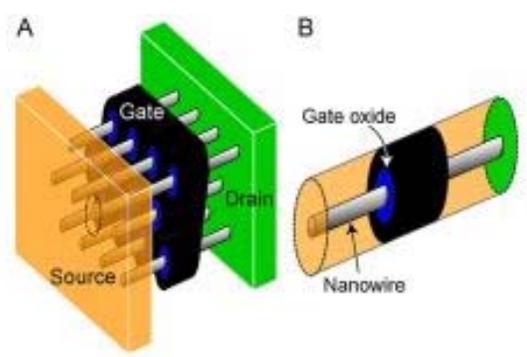
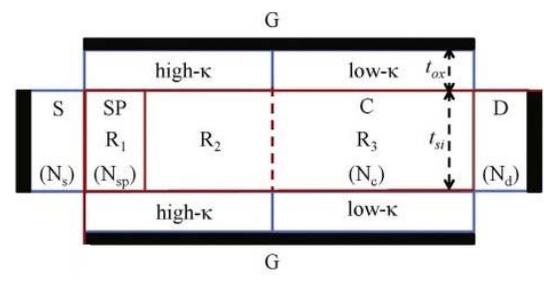

(e)                                           (f)

**Figure 1.7: Various TFET structures (a) Lateral Single Gate (b) Vertical (c) Double Gate (d) Pocketed (e) Nanowire (f) Hetero-gate**

- The TFET body can have doping profiles of p+ - i - n+, p+ - n - n+, p+ - p- - n+ etc. in source, channel and drains respectively. Sometimes narrow and highly doped regions called pockets are introduced in the channel to improve on-off ratio. Making the channel thin (ultra-thin body TFETs) can also improve on-off ratio. Some TFET designs also incorporate heterojunctions in the channel and use different semiconductor materials and alloys for bandgap engineering.
- Sometimes different insulators of different dielectric constants are incorporated in different parts of the gate dielectric to better control the Electric Field in the channel region and thus improve the subthreshold swing. These design of TFET is referred to as Hetero-gate (HG) TFET.

So this is apparent from this discussion that there can be numerous designs of TFET of varied complexities which aim to improve different parameters of the transistor such as, on current, on to off ratio and subthreshold swing.

## 1.8 Motivation

Tunnel FET is an emerging transistor technology. It is a subject of active research. Numerous designs of Tunnel FET are being proposed, experimented on and improved over time. It is crucial to develop proper tools for performance analysis of all these designs. As TFET is based on a different physics i.e. band to band tunneling they often require a different simulation technique than the well-established techniques used for MOSFETs. Most practical designs of TFET have feature sizes in the range of tens of nanometers or even less, which makes it imperative to include the effects of quantum mechanical phenomena like confinement and ballistic transport. The already existing models of TFETs have their limitations and simulation of TFETs often requires a compromise between time and accuracy. This motivated us to develop a numerical model of TFET which will be easy to use in simulations and also give good predictions and take into account the important quantum phenomena.

## 1.9 Thesis Layout

- In chapter 2 we will discuss the basic physics of TFETs and briefly describe existing physical models of TFET.

- We will develop our proposed model in chapter 3 where all the mathematical formulations will be presented in detail.
- In chapter 4 we will explain the step by step procedure required for implementation of our model in carrying out simulation of TFET.
- In chapter 5 we will present the results of performed simulations and discuss about their implications.
- Finally, chapter 6 we will end with some conclusive remarks and talk about prospects of future work.

# Chapter 2
# Theoretical Overview

In this chapter we present the basic physics required to explain band-to-band tunneling which is the fundamental principle of operation of TFETs. Then we introduce various models of BTBT, their limitations and applicability. Finally we point out the necessity of a new 2D quantum mechanical model and discuss the objectives of our thesis.

## 2.1 Quantum Mechanical Tunneling

According to classical mechanics, if a particle such as an electron is incident on a barrier with kinetic energy lower than the potential energy of the barrier, it can never pass through that barrier. But quantum mechanics suggests that an electron with energy lower than the barrier potential energy has a finite probability of passing through the barrier because of its wave nature [36] as shown in Figure 2.1. This phenomenon is known as quantum mechanical tunneling.

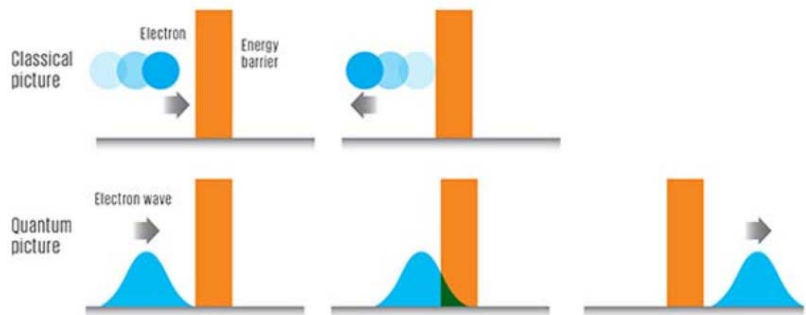

**Figure 2.1: Electron incident on a barrier; classical picture (above) where it can never pass vs quantum mechanical picture (below) where it may pass with finite probability**

The electron travels as an evanescent wave in the barrier with an imaginary wave vector ($k$) which means that the amplitude of the wave function decays exponentially with distance and so does the probability of finding the electron in the barrier (Figure 2.2). After crossing the barrier the wave function again takes the form of a traveling wave but has a reduced amplitude.

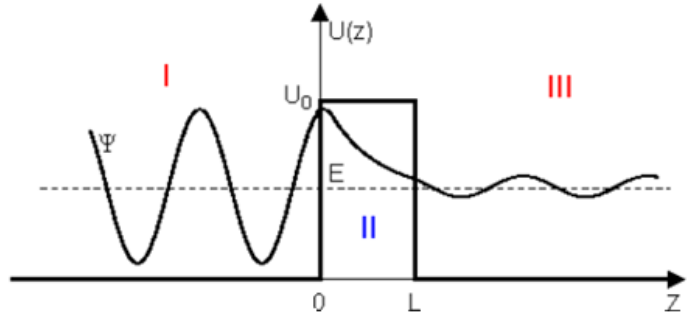

**Figure 2.2: Wave function ψ for a tunneling electron, I-incident; II-barrier; III-transmitted**

### 2.1.1 Transmission Coefficient

The tunneling probability of an electron is given by the Transmission Coefficient which depends on the energy of the electron, barrier potential height, length and also on the mass of the electron. The transmission coefficient for a constant barrier height of $U_0$ and length $L$, is given by the approximate formula [36]

$$T(E) \approx \exp(-2kL) \qquad (2.1)$$

Where, $k$ is the imaginary wave vector in the barrier, given by

$$k = \sqrt{\frac{2m(U_0 - E)}{\hbar^2}} \qquad (2.2)$$

Here, $m$ is the mass of the electron and $\hbar$ is the reduced Plank's constant.

## 2.2 The WKB Approximation

If the barrier potential height varies arbitrarily over its length, like that shown in Figure 2.3, it becomes difficult to obtain an exact expression for transmission coefficient for a particle tunneling through this barrier, from simple analysis. Often we don't need an exact expression but rather an approximate one like equation (2.1). The Wentzel–Kramers–Brillouin (WKB) approximation aids us in this case [9] and gives a very useful tool for calculation of tunneling current.

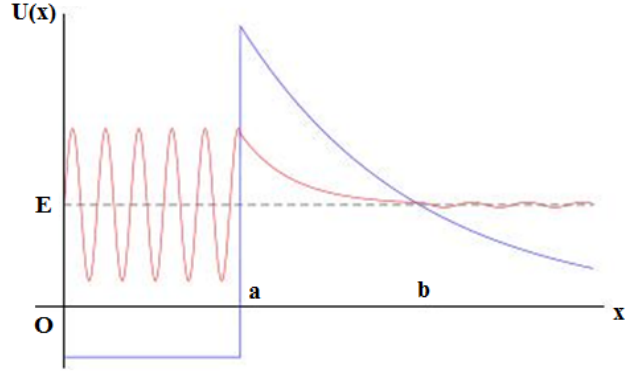

**Figure 2.3: A barrier of varying potential energy and the wave function of tunneling electron in different regions**

If an electron of energy $E$ travels in the $x$ direction when it faces a varying potential barrier $U(x)$, then WKB approximation gives the following approximate value of the transmission coefficient, denoted by $T_{WKB}(E)$.

$$T_{WKB}(E) = \exp\left(-2 \int_a^b |k(x)| dx\right) \tag{2.3}$$

Where, the wave vector $k$ is given by

$$k = \sqrt{\frac{2m(E - U(x))}{\hbar^2}} \tag{2.4}$$

Here, $m$ is the mass of the electron or any other particle under consideration, which is replaced by the effective mass in case of a quasi-particle in a crystal.

### 2.2.1 Physical Explanation of the WKB Approximation for Tunneling

The WKB approximation for tunneling electron has a sound mathematical formulation, but to get a physical sense of how may we get this expression, consider dividing the barrier in infinitesimally small slices of constant height, and then applying equation (2.1) to get the transmission coefficient for each slice. By ignoring internal reflections and multiplying all the transmission probabilities we end up with the overall transmission coefficient given by equation (2.3). This formulation has the physical implication that equation (2.3) will only work well when the barrier is much higher than the energy of the electron so that we can neglect internal reflections of the wave vector.

### 2.2.2 Limitations of WKB Approximation

The WKB approximation gives a simple but good approximation of the transmission coefficient. Notice however that, in equation (2.3) to get the transmission coefficient we need to integrate over a line (in this case along the x axis) which represents the tunneling path. This is a trivial task if the potential distribution is one dimensional (1D) because there is only one possible tunneling path and we do not need to make any predictions. However, for a 2D or a 3D potential distribution we need to predict the tunneling path before applying WKB approximation which can be difficult.

## 2.3 The Physics of Tunnel FET: Band to Band Tunneling

As we have already mentioned, band to band tunneling (also called Zener tunneling or interband tunneling) is the process in which electron tunnels from one band to another which involves passing through the bandgap [9]. Although the actual wave function of an electron in a crystal is quite complicated; due to Bloch theorem it can be expressed as an envelope function modulated by a function having the periodicity of the crystal called Bloch function [37]. So, for an electron in the bands, the envelope function takes the role of the wave function of a free electron and the wave vector is replaced by the crystal momentum which is also denoted by $k$ for simplicity. In the bandgap the crystal momentum $k$ is imaginary just like the wave vector for a free electron is imaginary in the barrier as discussed in the previous section. So, often the process of band to band tunneling can be approximated by the process of electron passing through a barrier (which is situated in the bandgap) as an evanescent wave. However, unlike the case of free electron, the dispersion relationship i.e. the $E$ vs. $k$ relationship is not parabolic in the bandgap. To calculate the transmission coefficient for BTBT, we would need to know the dispersion relationship in the bandgap. To simplify the analogy between free electron and the BTBT electron often an approximation of the effective mass called the tunneling effective mass is used. Then the effective mass transitions from hole effective mass to electron effective mass during tunneling.

If we can calculate the imaginary wave vector in the bandgap (in this case actually crystal momentum, but to simplify the analogy we will refer to it as the wave vector) and the tunneling path, we may invoke WKB approximation to calculate the transmission coefficient. The tunneling current can be calculated from the transmission coefficient.

## 2.3.1 Types of Band to Band Tunneling

Based on whether crystal momentum of the electron is conserved during the tunneling process, band to band tunneling can be classified into 2 types [9].

### 2.3.1.1 Direct BTBT

In direct BTBT the momentum $k$ is conserved. This is possible if the conduction band minimum and the valence band maximum occur at the same value of $k$. This condition is satisfied by direct bandgap materials such as GaAs and also by some indirect bandgap material like Ge at high electric field. Direct BTBT is the dominant BTBT process is direct bandgap semiconductors.

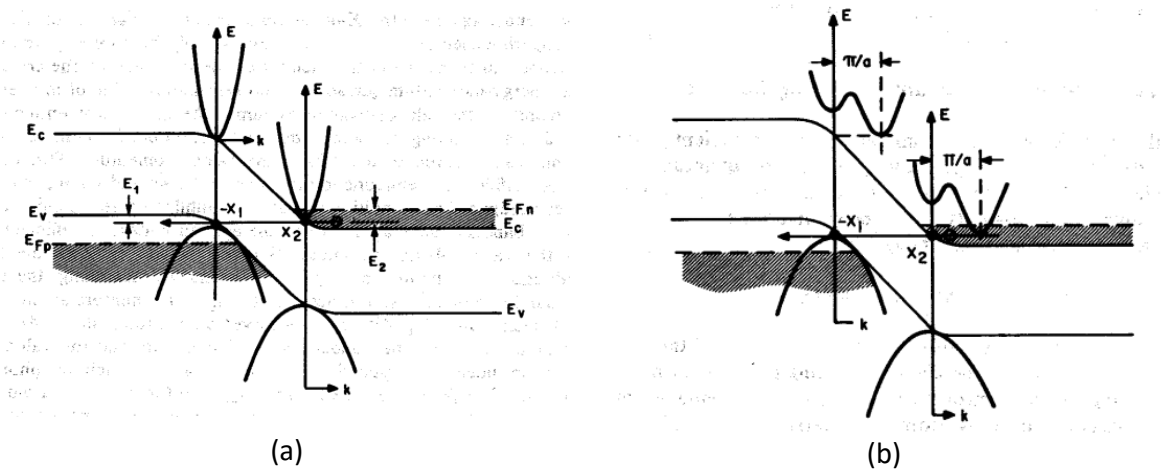

**Figure 2.4: Types of BTBT (a) Direct (b) Indirect**

### 2.3.1.2 Indirect BTBT

In indirect BTBT momentum $k$ is not conserved because the conduction band minimum and the valence band maximum do not occur at the same value of $k$. So, the difference in momentum must be supplied by scattering agents such as traps or phonons. This is the dominant BTBT process in indirect bandgap materials such as Si. Depending on the type of scattering agent participating BTBT process indirect BTBT can be of two types, namely phonon assisted and trap assisted. In case of phonon assisted tunneling both energy and momentum must be conserved in the scattering process.

## 2.4 Modeling Band to Band Tunneling

Modeling BTBT requires a different strategy than the general semi-classical modeling approach where it is assumed that electrons and holes are quasi-particles and they stay in their respective bands except for the thermal generation-recombination process.

Modeling BTBT requires proper treatment of dispersion relations in the bandgap for evanescent wave vector. This is different from the effective mass approximation which applies to a single band. Band to band tunneling direction usually does not follow any preferred direction unless the device is very thin (like nanowire TFETs). So any assumed direction of tunneling may lead to inconsistent results. Also, the electron hole pair generated by BTBT appear at two different spatial points, unlike the thermal generation process.

Most conventional methods of semiconductor device simulation are semi-classical in nature and cannot properly account for quantum mechanical phenomena like quantum confinement, ballistic transport and intraband tunneling. The models developed for quantum mechanical simulation of MOSFET based on the single band NEGF or uncoupled mode space approach deal with one band and one type of carriers, and cannot be used for TFET simulation without proper elaboration. The multi-band models remedy this problem but add additional numerical overhead.

Quantum atomistic simulation models, based on the first principal seem to produce the most reliable results. But such approach requires considerable amount of time and computational resources than semi-empirical models.

### 2.4.1 Band to Band Tunneling Models

The Models of BTBT can be classified into 2 categories, namely semi-classical and quantum mechanical.

#### 2.4.1.1 Semi-classical Models:

Semi-classically BTBT can be modelled as a generation-recombination process of electron hole pairs. Semi-classical models use WKB approximation along with macroscopic electric field/potential to calculate generation rates. Classical drift-diffusion equations are used to model the transport of generated carriers. The bulk density of states with Fermi-Dirac or Maxwell-Boltzmann are used to calculate the carrier concentrations. Quantum mechanical effects like confinement and intra-band tunneling are ignored.

**Table 2.1: Models of Band to Band Tunneling**

| Semi-classical Models | Quantum Mechanical Models |
|---|---|
| ❖Local Models<br>➤ Kane model<br>➤ Hurkx model<br>➤ Schenk model<br>❖Non-local Models<br>➤ Static Path<br>➤ Dynamic Path | ❖Effective Mass Approximation<br>➤ Non-local with confinement<br>➤ Phonon assisted<br>❖Long Channel Device<br>➤ Monte-Carlo<br>❖Short Channel Device<br>➤ Atomistic k.p<br>➤ Tight binding |

The advantage of semi-classical models of BTBT is that they can be seamlessly integrated into classical models of semiconductor by treating BTBT as a generation process along with other generation/recombination processes.

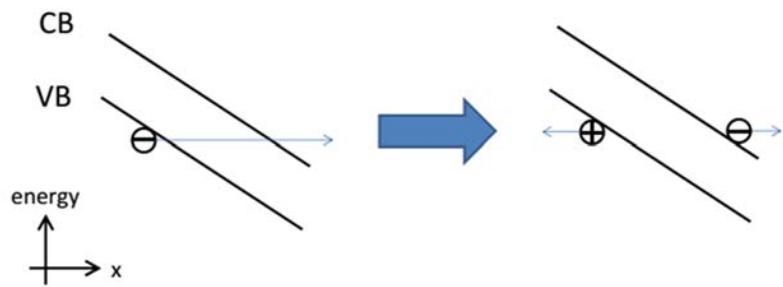

**Figure 2.5: In semiclassical BTBT models tunneling of electron is modeled as generation of electron hole pairs in respective bands**

2.4.1.1.1 Local Models:

Local models consider BTBT as a point generation process, i.e. electrons and holes generate at the same point [38]. They make the assumption that the electric field is constant over the tunneling length to simplify calculation of transmission coefficient. Due to this assumption BTBT carrier generation rate depends solely on the magnitude of electric field. The simplicity and ease of use of local models make them suitable for developing analytical models but not so much for accurate simulation of TFET.

Example: The well-known Kane's model of BTBT [21]

$$G^{BTBT} = A \frac{F^\gamma}{\sqrt{E_g}} \exp\left(-B \frac{E_g^{\frac{3}{2}}}{F}\right) \quad (2.5)$$

Here, $F$ is the electric field. $A, B, \gamma$ are model parameters.

### 2.4.1.1.2 Non-local Models

Unlike local models which assume a constant electric-field and thus constant slope of bands in the tunneling direction, non-local models use actual shapes of the bands throughout the device to calculate transmission coefficient [38]. They also consider generation of electron and hole to take place at the edges of the tunneling path rather than at the same point.

Non-local models use WKB approximation to calculate the transmission coefficient for which a 1D tunneling path needs to be assumed

- Static path models use a predefined tunneling path; for example, a straight line from source to drain
- Dynamic path models try to predict the most probable tunneling path from the direction of electric field or curvature of the bands

Non-local models give better results in simulations than local models, if the assumed tunneling path is correct. So they are widely used in semi-classical simulation of TFETs.

### 2.4.1.2 Quantum Mechanical Models

Most quantum mechanical models use multi-band approach to approximate the dispersion relationship in the bandgap. However there are some models that can be invoked under single band effective mass approximation.

Monte Carlo simulations use Boltzmann transport with Fermi's golden rule. They are more appropriate for devices where channel is long and scattering is high.

k.p and tight binding models produce highly accurate results for nanometer devices. Here, often ballistic or quasi-ballistic transport is assumed and NEGF formalism is invoked to solve the Schrodinger's equation.

A recent work by Vandenberghe et al. [39] uses spectral function to approximate phonon assisted tunneling under closed boundary conditions.

Quantum mechanical effects like confinement and intra-band tunneling are inherently accounted for in these models. It is also possible to incorporate some of the quantum phenomena in semi-classical models to increase their accuracy.

### 2.4.2 Limitations of BTBT Models used in TFET simulation

Non-Local Semi-classical models use WKB approximation in which a 1D tunneling path needs to be assumed. This only works perfectly when the potential profile is perfectly 1D. But In reality most of the devices show pronounced 2D potential distribution. So, prediction of proper tunneling path can fail and sometimes parasitic tunneling paths are overlooked [39]. This is even more problematic if the channel contains heterojunctions, i.e. abrupt change in band profile.

After some modifications, the non-local models can take into account quantum mechanical effects like confinement [40] which is prominent in most TFETs due to their low dimensionality. But still this can lead to wrong predictions. Quantum phenomena like ballistic transport and intra-band tunneling cannot be incorporated easily and mesoscopic factors like resonant tunneling, or sub-bands in super-lattice structures cannot be included.

The quantum mechanical models of band to band tunneling based on effective mass approximation cannot properly account for ballistic transport. There are some models based on 1D NEGF [41] which fail take into account the actual tunneling paths in thick body devices.

Multiband quantum mechanical methods like k.p and tight-binding are very precise but they are also significantly complex and require more computational resources. They also consume a lot of time which can be problematic in iterative design process [42].

### 2.5 The Necessity of a 2D Model

It is common for devices to have a 2D geometry, the same is true for TFETs. Potential varies along device thickness and so does the band profile. So tunneling does not occur in any single direction.

A typical potential distribution and BTBT generation rate in the channel region of a TFET is shown in Figure 2.6.The region right underneath the gate have the highest spatial curvature in electric potential (Figure 2.6(a))., but this is where most of the BTBT occurs (Figure 2.6(b)). A 1D model often fails to predict proper tunneling path, and gives less generation rate (and consequently less current) than actual value. This problem is more severe in some electron-hole bilayer [43] or L shaped TFETs where unexpected parasitic tunneling paths gets ignored, due to the highly skewed potential distribution in those devices.

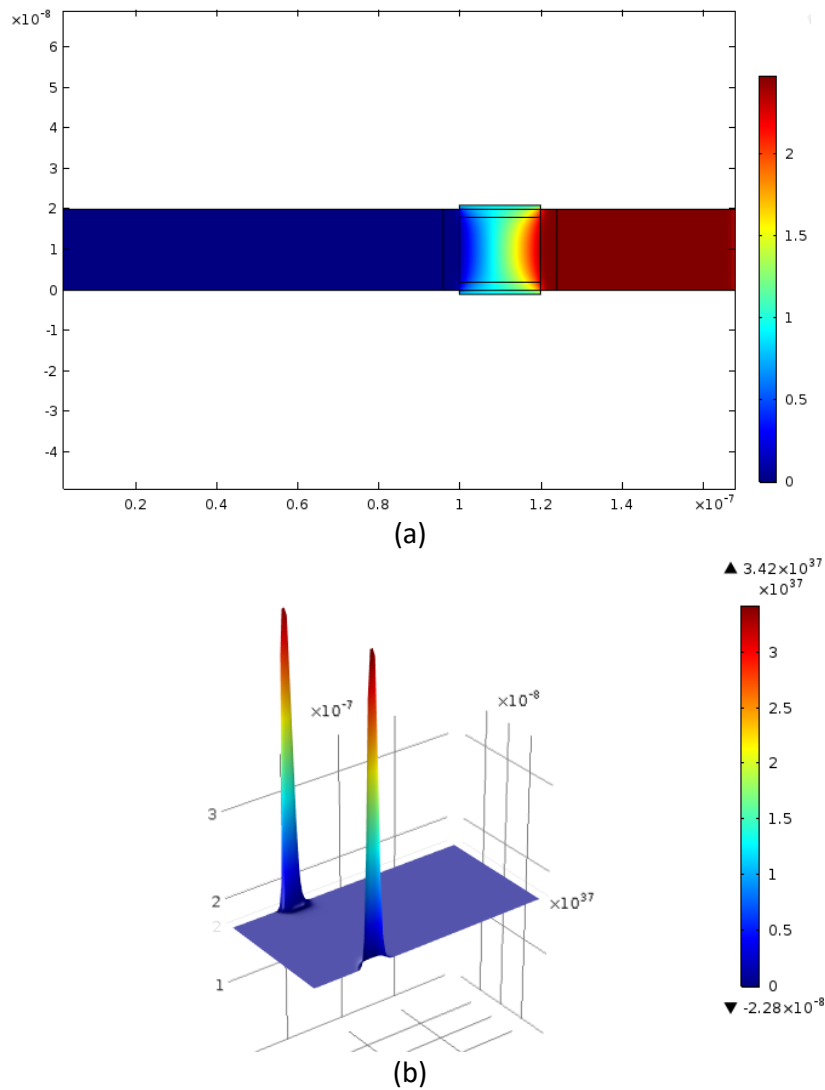

**Figure 2.6: (a) Electric potential profile in a typical lateral DG TFET (b) BTBT generation rate in the channel region, which is highest just beneath the gate**

The dynamic non-local path models try to account for the 2D nature of the band profile by trying to predict the most probable tunneling paths [38] and applying WKB approximation along that path. Then the actual current is found by integrating the currents along all paths. As shown in Figure 2.7, it is like slicing up a 2D sheet into thin 1D strips to measure current flow. The resistances of strips can be considered to be in parallel only if their resistances are equal. We do not know how to slice it, and even if we do, it is more complicated than simply putting them in parallel.

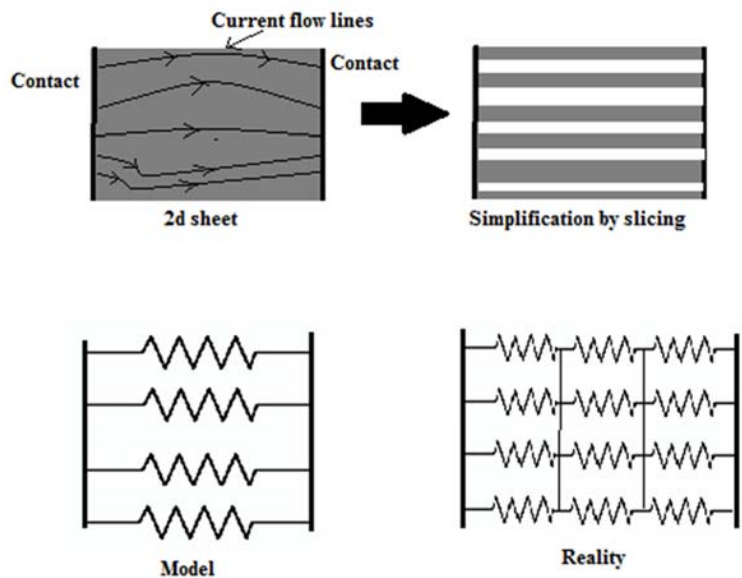

**Figure 2.7: Calculating current by making 1D slices (narrow tunneling paths) of a 2D device: an analogy with resistance calculation**

For some devices like that in Figure 2.8 the structure [44] is neither lateral nor vertical and the direction of tunneling is highly unpredictable. Different tunneling direction can lead to catastrophically different results when calculating current. Also, tunneling path prediction based on electric field does not work well when heterojunctions are involved.

Again in the realm of ballistic transport, one cannot simply put different current paths in parallel because of interference between different parts of the wave function. So, the 2D geometry of the TFET needs proper quantum mechanical treatment.

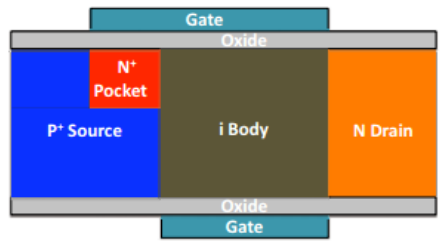

**Figure 2.8: A Heterojunction Vertical TFET structure [44]: in this device current results from a mixture of BTBT in lateral and vertical directions**

## 2.6 Thesis Objective

In light of the discussion on the limitations of TFET models we want to develop a quantum mechanical model of TFET based on 2D NEGF formalism that

- ➢ Addresses the shortcomings of semi-classical models, like the requirement for a 1D tunneling path
- ➢ Gives exact treatment of 2D potential distribution which is common in TFETs
- ➢ Extends the single band effective mass approximation to account for BTBT
- ➢ Can be integrated with already existing single band quantum ballistic model
- ➢ Takes into account the basic quantum mechanical effects like confinement and intra-band tunneling
- ➢ Faster than multi-band quantum simulations like k.p or tight-binding

That is we want a compromise between semi-classical and multi-band quantum simulations.

## 2.7 Chapter Summary

In this chapter we have discussed about basic TFET physics and models. In light of the limitations of the existing models we have stated the objective of this thesis work to qualitatively point out the nature of the model we wish to develop.

# Chapter 3
# Development of the Model

In this chapter we develop our model based on step by step mathematical formulation. First we will simplify the 3D Schrodinger's Equation for 2D potential distribution. Then we formulate an approximate dispersion relation in the bandgap. Next, we determine the effective potential and mass in different regions to formulate the final form of Schrodinger's equation.

After we have the Schrodinger's equation we move on to the numerical analysis of the model. We discretize the Schrodinger's equation and determine the Hamiltonian and self-energy matrices. Then we invoke the 2D NEGF formalism to determine the carrier concentrations and current. We will simplify the results of NEGF formalism for faster simulation. Finally we will discuss how our model may be implemented in simulations of TFET.

## 3.1 Simplification of Schrodinger Equation for 2D Potential Distribution

The Schrodinger's equation is written as-

$$\hat{H}\Psi = E\Psi \tag{3.1}$$

Where, $\hat{H}$ is the Hamiltonian operator and $E$ is the energy of a state. In valence band, under effective mass approximation and for a 3D potential distribution, single band Hamiltonian is

$$\hat{H} \equiv \frac{\hbar^2}{2m_v^*}\nabla^2 + E_v(x,y,z) \tag{3.2}$$

Where, $m_v^*$ is the effective mass in valence band (i.e. hole effective mass)

$$\nabla^2 \equiv \frac{\partial^2}{\partial x^2} + \frac{\partial^2}{\partial y^2} + \frac{\partial^2}{\partial z^2} \tag{3.3}$$

And, the valence band profile is given by

$$E_v(x,y,z) = E_{v_0} + U_E(x,y,z) \tag{3.4}$$

Where, $E_{v_0}$ is the valence band maxima and $U_E$ is the external applied electric potential. If $U_E$ is invariant in the z direction, i.e. if the potential distribution is 2D, then $U_E$ and thus $E_v$ is only dependent on $x$ and $y$ coordinates.

So, simplifying-

$$E_v(x,y) = E_{v_0} + U_E(x,y) \tag{3.5}$$

So, the Schrodinger's equation given in (3.1) becomes

$$\left[\frac{\hbar^2}{2m_v^*}\left(\frac{\partial^2}{\partial x^2} + \frac{\partial^2}{\partial y^2} + \frac{\partial^2}{\partial z^2}\right) + E_v(x,y)\right]\Psi(x,y,z) = E\Psi(x,y,z) \tag{3.6}$$

Here, $\Psi$ is actually the envelope function in the valence band, but we will keep calling it the wavefunction for simplicity. Now we simplify this equation based on the fact that $E_v$ is invariant along the z axis. To do so, let us consider the 1D equation in the z direction,

$$\left[\frac{\hbar^2}{2m_v^*}\frac{d^2}{dz^2} + E_v(x,y)\right]\phi_{k_z}(z;x,y) = E_{v,\text{sub},k_z}(x,y)\phi_{k_z}(z;x,y) \tag{3.7}$$

$$\Rightarrow \hat{H}_z\phi_{k_z}(z;x,y) = E_{v,\text{sub},k_z}(x,y)\phi_{k_z}(z;x,y) \tag{3.8}$$

Where $E_{v,\text{sub},k_z}(x,y)$ may be called the sub-band energy for a particular value of $k_z$, i.e. the z component of wave vector (crystal momentum) $\vec{k}$. The reason for isolating $k_z$ will be apparent shortly. Here $\hat{H}_z$ is a Hermitian operator, so the eigenvectors of $\hat{H}_z$ i.e. $\phi_{k_z}(z;x,y)$ form a complete set of orthonormal basis vectors. Which enables us to write $\Psi(x,y,z)$ as the linear combination-

$$\Psi(x,y,z) = \sum_{k_z}\psi_{k_z}(x,y)\phi_{k_z}(z;x,y) \tag{3.9}$$

Now, let us go back to equation (3.7) which can be written as

$$\frac{d^2\phi_{k_z}(z;x,y)}{dz^2} + \frac{2m_v^*[E_v(x,y) - E_{v,\text{sub},k_z}(x,y)]}{\hbar^2}\phi_{k_z}(z;x,y) = 0 \tag{3.10}$$

The differential equation can be solved as

$$\frac{d^2\phi_{k_z}(z;x,y)}{dz^2} + k_z^2\phi_{k_z}(z;x,y) = 0 \tag{3.11}$$

Where, $k_z$ is given by

$$k_z = \sqrt{\frac{2m_v^*[E_v(x,y) - E_{v,\text{sub},k_z}(x,y)]}{\hbar^2}} \tag{3.12}$$

Solving for $\phi_{k_z}(z; x, y)$ and applying boundary conditions, we find discrete values of $k_z$ given by

$$k_z = \frac{n\pi}{L_z} \tag{3.13}$$

Where $L_z$ is the device dimension along the $z$ direction, i.e. the direction along which the potential is invariant.

Now, since the device is assumed to be wide, and there is no confinement along the z direction, the values of $k_z$ can be taken to be quasi-continuous. Then the number of states lying between $k_z$ and $k_z + dk_z$ can be written as:

$$\rho(k_z)dk_z = \frac{L_z}{\pi}dk_z \tag{3.14}$$

$$\text{i.e. } \rho(k_z) = \frac{L_z}{\pi} \tag{3.15}$$

Now, returning to equation (3.6) and putting the value from equation (3.9) we get,

$$\left[\frac{\hbar^2}{2m_v^*}\left(\frac{\partial^2}{\partial x^2} + \frac{\partial^2}{\partial y^2} + \frac{\partial^2}{\partial z^2}\right) + E_v(x,y)\right]\sum_{k_z'}\psi_{k_z'}(x,y)\phi_{k_z'}(z;x,y) = E\sum_{k_z'}\psi_{k_z'}(x,y)\phi_{k_z'}(z;x,y) \tag{3.16}$$

$$\Rightarrow \sum_{k_z'}\frac{\hbar^2}{2m_v^*}\left[\left(\frac{\partial^2}{\partial x^2} + \frac{\partial^2}{\partial y^2}\right)\psi_{k_z'}(x,y)\right]\phi_{k_z'}(z;x,y)$$
$$+ \sum_{k_z'}\left[\frac{\hbar^2}{2m_v^*}\frac{d^2\phi_{k_z'}(z;x,y)}{dz^2}\right.$$
$$\left. + E_v(x,y)\phi_{k_z'}(z;x,y)\right]\psi_{k_z'}(x,y)$$
$$= E\sum_{k_z'}\psi_{k_z'}(x,y)\phi_{k_z'}(z;x,y) \tag{3.17}$$

Putting value from equation (3.7) in this equation, we get

$$\sum_{k_z'} \frac{\hbar^2}{2m_v^*} \phi_{k_z'}(z;x,y) \nabla_{xy}^2 \psi_{k_z'}(x,y)$$

$$+ \sum_{k_z'} E_{v,\text{sub},k_z'}(x,y) \phi_{k_z'}(z;x,y) \psi_{k_z'}(x,y) \qquad (3.18)$$

$$= E \sum_{k_z'} \psi_{k_z'}(x,y) \phi_{k_z'}(z;x,y)$$

Where, we have used the notation for the operator

$$\nabla_{xy}^2 \equiv \frac{\partial^2}{\partial x^2} + \frac{\partial^2}{\partial y^2} \qquad (3.19)$$

Now, multiplying both sides by $\phi_{k_z}(z;x,y)$ and integrating in the z direction we get,

$$\sum_{k_z'} \frac{\hbar^2}{2m_v^*} \nabla_{xy}^2 \psi_{k_z'}(x,y) \int \phi_{k_z}(z;x,y) \phi_{k_z'}(z;x,y) \, dz$$

$$+ \sum_{k_z'} E_{v,\text{sub},k_z'}(x,y) \psi_{k_z'}(x,y) \int \phi_{k_z}(z;x,y) \phi_{k_z'}(z;x,y) \, dz \qquad (3.20)$$

$$= E \sum_{k_z'} \psi_{k_z'}(x,y) \int \phi_{k_z}(z;x,y) \phi_{k_z'}(z;x,y) \, dz$$

Due to the orthonormality property of $\phi_{k_z}$ vectors,

$$\int \phi_{k_z}(z;x,y) \phi_{k_z'}(z;x,y) \, dz = \delta(k_z - k_z') \qquad (3.21)$$

Where $\delta(x)$ is the Kronecker delta function.

So equation (3.20) yields, the simplified form of Schrodinger's equation

$$\frac{\hbar^2}{2m_v^*} \nabla_{xy}^2 \psi_{k_z}(x,y) + E_{v,\text{sub},k_z}(x,y) \psi_{k_z}(x,y) = E \psi_{k_z}(x,y) \qquad (3.22)$$

Where, from equation (3.12)

$$E_{v,\text{sub},k_z}(x,y) = E_v(x,y) - \frac{\hbar^2 k_z^2}{2m_v^*} \qquad (3.23)$$

Similarly, in the conduction band, the simplified form of Schrodinger's equation

$$-\frac{\hbar^2}{2m_c^*}\nabla_{xy}^2\psi_{k_z}(x,y) + E_{c,sub,k_z}(x,y)\psi_{k_z}(x,y) = E\psi_{k_z}(x,y) \quad (3.24)$$

Where, $E_{c,sub,k_z}$ is the sub-band energy for a particular value of $k_z$ and given by

$$E_{c,sub,k_z}(x,y) = E_c(x,y) + \frac{\hbar^2 k_z^2}{2m_c^*} \quad (3.25)$$

Where, $m_c^*$ is the effective mass in conduction band (i.e. electron effective mass) and $E_c(x,y)$ is the conduction band profile.

So, we have reduced the 3D Schrodinger's equation to 2D form, in both conduction and valence band. Figure 3.1 shows the conduction and valence band profiles and corresponding sub-band profiles for two different values of $k_z$. Notice that the sub-bands move up in the conduction band and down in the valence band with increasing $k_z$.

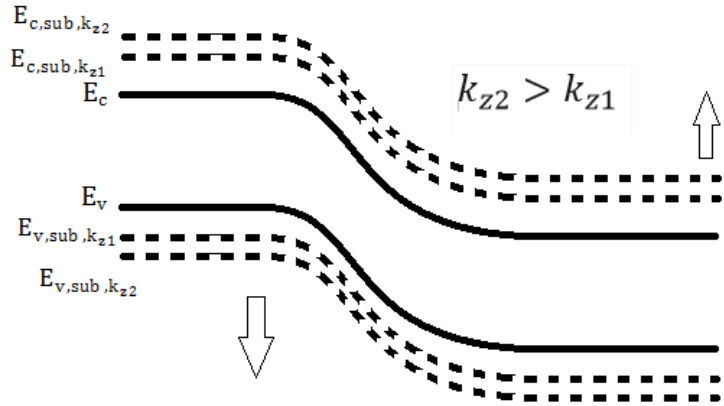

Figure 3.1: The bands (solid) and the sub-bands (dashed)

## 3.2 Conservation Laws in Direct Band to Band Tunneling

In this model we will consider only direct band to band tunneling, which is an elastic process [9], so the total energy is conserved for a tunneling electron i.e.

$$\underbrace{\widehat{E_{vb}}}_{Energy\ in\ valence\ band} = \underbrace{\widehat{E_{cb}}}_{Energy\ in\ conduction\ band} = E \quad (3.26)$$

By approximating BTBT by a tunneling process through a barrier, Schrodinger's equation for the tunneling electron can be written as

$$\left[\frac{\hbar^2}{2m_v^*}\left(\frac{\partial^2}{\partial x^2}+\frac{\partial^2}{\partial y^2}+\frac{\partial^2}{\partial z^2}\right)+U_{barrier}(x,y,z)\right]\Psi(x,y,z)=E\Psi(x,y,z) \quad (3.27)$$

Where, $U_{barrier}(x,y,z)$ is the virtual potential barrier in the bandgap. If external electrical potential is invariant along z direction, it is reasonable to assume that $U_{barrier}$ will be invariant along z direction as well. That is:

$$\frac{\partial U_{barrier}}{\partial z}=0 \quad (3.28)$$

By solving the Schrodinger's equation it can be shown that the z component of the wave vector (crystal momentum) $\vec{k}$ i.e. $k_z$ will remain unchanged during tunneling.

But we can get an explanation for this phenomenon even from a semi-classical point of view. According to the classical definition of force-

$$F_z=-\frac{\partial U_{barrier}}{\partial z}=0 \quad (3.29)$$

Where, $F_z$ is z component of the force applied on the electron during tunneling. It follows from Newton's law of motion:

$$\frac{\partial p_z}{\partial t}=0 \quad (3.30)$$

$$\Rightarrow p_z=\hbar k_z=constant \quad (3.31)$$

$$So, k_z=constant \quad (3.32)$$

So the sub-bands participating in the tunneling process must correspond to the same value of $k_z$.

## 3.3 Dispersion Relationship in the Bandgap

To estimate the barrier that the electron faces in the bandgap we need an approximate dispersion relationship in the bandgap. The wave vector (crystal momentum) $\vec{k}$ has imaginary component in bandgap which indicates an evanescent wave function whose amplitude decays over distance.

An exact treatment of dispersion relationship would require analyzing the band structure of the semiconductor material by analyzing the crystal structure at the atomistic level. However, just like parabolic band approximation in the conduction or valence band simplifies the dispersion relationship in the corresponding band, simplification of the dispersion relation can be made in the bandgap as well. Several approximate models exists for this purpose.

We will use the simplified 2 band approximation by Kane and Franz [45, 46]. In this model, the dispersion relation in the bandgap is given by

$$|\vec{k}|^2 = \frac{2m_t^*}{\hbar^2} \frac{(E - E_c)(E - E_v)}{E_g} \tag{3.33}$$

Where, $m_t^*$ is the tunneling effective mass and $E_g$ is the bandgap. We will address the variation of tunneling effective mass with energy in a latter section. For now, let us simplify the relation given by the 2 band approximation to better suit our purpose

$$\frac{1}{|\vec{k}|^2} = \frac{\hbar^2}{2m_t^*} \frac{E_c - E_v}{(E - E_c)(E - E_v)} = \frac{\hbar^2}{2m_t^*} \left[ \frac{1}{E - E_c} + \frac{1}{E_v - E} \right] \tag{3.34}$$

Now, as we know, the dispersion relationship in conduction band is given by the effective mass approximation as:

$$E = E_c + \frac{\hbar^2 |\vec{k_c}|^2}{2m_c^*} \Rightarrow \frac{1}{|\vec{k_c}|^2} = \frac{\hbar^2}{2m_c^*} \frac{1}{E - E_c} \tag{3.35}$$

Similarly, from the dispersion relationship in valence band

$$E = E_v - \frac{\hbar^2 |\vec{k_v}|^2}{2m_v^*} \Rightarrow \frac{1}{|\vec{k_v}|^2} = \frac{\hbar^2}{2m_v^*} \frac{1}{E_v - E} \tag{3.36}$$

Here $\vec{k_c}$ and $\vec{k_v}$ are single band approximations of the wave vector $\vec{k}$ in the conduction and valence bands respectively.

Near the conduction band $m_t^* \approx m_c^*$ and near the valence band $m_t^* \approx m_v^*$ so, combining equations (3.34), (3.35) and (3.36) we can write, the 2 band dispersion relationship as,

$$\frac{1}{|\vec{k}|^2} \approx \frac{1}{|\vec{k_c}|^2} + \frac{1}{|\vec{k_v}|^2} \tag{3.37}$$

So, basically the 2 band model indicates that we should take the harmonic mean of the squared magnitude of the single band approximations of the wave vectors to ensure that the dispersion relationship is electron-like near the conduction band (when $|\vec{k_c}|^2$ $is\ very\ small$) and hole-like near the valence band (when $|\vec{k_v}|^2$ $is\ very\ small$). This allows a smooth transition of dispersion relation from conduction to valence band as shown in Figure 3.2. There is a subtle issue of the tunneling effective mass which shapes the curvature of E-k curve and it will addressed in the next section.

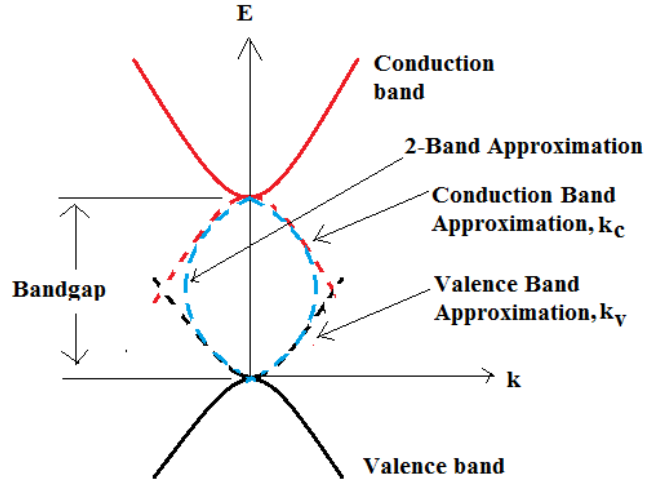

**Figure 3.2: The dispersion relationship given by single band and 2 band models; dashed lines indicate imaginary wave vector**

Now to simplify the dispersion relation for the case of 2D potential, we use the result of the previous section, which shows that the $z$ component of $\vec{k}$ i.e. $k_z$ remains constant during tunneling. Writing

$$|\vec{k}|^2 = k_x^2 + k_y^2 + k_z^2 = k_{xy}^2 + k_z^2 \tag{3.38}$$

Where we have lumped the squared values $x$ and $y$ components of $\vec{k}$ in $k_{xy}^2$. Since the $k_z$ component remains unchanged during tunneling we leave it out of the dispersion relation to get a better approximation-

$$\frac{1}{k_{xy}^2} = \frac{1}{k_{c,xy}^2} + \frac{1}{k_{v,xy}^2} \tag{3.39}$$

Where, subscripts 'c' and 'v' denote single band approximations in conduction and valence bands respectively. Physically this relationship indicates an increase in effective bandgap with increase in $k_z$. This is to be expected since the electron have to travel a longer forbidden range of energy than the bandgap (Figure 3.4), i.e. the gap between the valence band maximum and conduction band minimum.

Now, to find the single band approximation in valence band

$$E = E_v(x,y) - \frac{\hbar^2 k_v^2(x,y)}{2m_v^*} = E_v(x,y) - \frac{\hbar^2 k_z^2}{2m_v^*} - \frac{\hbar^2 k_{v,xy}^2(x,y)}{2m_v^*} \quad (3.40)$$
$$= E_{v,sub,k_z}(x,y) - \frac{\hbar^2 k_{v,xy}^2(x,y)}{2m_v^*}$$

$$So, k_{v,xy}^2(x,y) = \frac{2m_v^*[E_{v,sub,k_z}(x,y) - E]}{\hbar^2} \quad (3.41)$$

Similarly, in conduction band,

$$k_{c,xy}^2(x,y) = \frac{2m_c^*[E - E_{c,sub,k_z}(x,y)]}{\hbar^2} \quad (3.42)$$

Now the wave vector in the bandgap is found from equation () as:

$$k_{xy}^2(x,y) = \frac{k_{v,xy}^2(x,y) k_{c,xy}^2(x,y)}{k_{c,xy}^2(x,y) + k_{v,xy}^2(x,y)} \quad (3.43)$$

We want to find the effective barrier seen by the tunneling electron in the bandgap. The dispersion relationship of the tunneling electron in the bandgap is then given by

$$E = U_{b,k_z}(x,y) + \frac{\hbar^2 k_{xy}^2(x,y)}{2m_t^*(E)} \quad (3.44)$$

$$So, U_{b,k_z}(x,y) = E - \frac{\hbar^2 k_{xy}^2(x,y)}{2m_t^*(E)} \quad (3.45)$$

Where $U_{b,kz}(x,y)$ is the effective barrier seen by the electron for a particular value of $k_z$. Here it should be mentioned that $k_{xy}(x,y)$ will be imaginary in the bandgap, so the barrier will be higher than E, as expected. Also, as we can see from Figure 3.3(b), barrier $U_{b,kz}(x,y)$ increases both in length and height with increase in $k_z$ for a particular value of $E$, so tunneling becomes less and less probable for higher values of $k_z$.

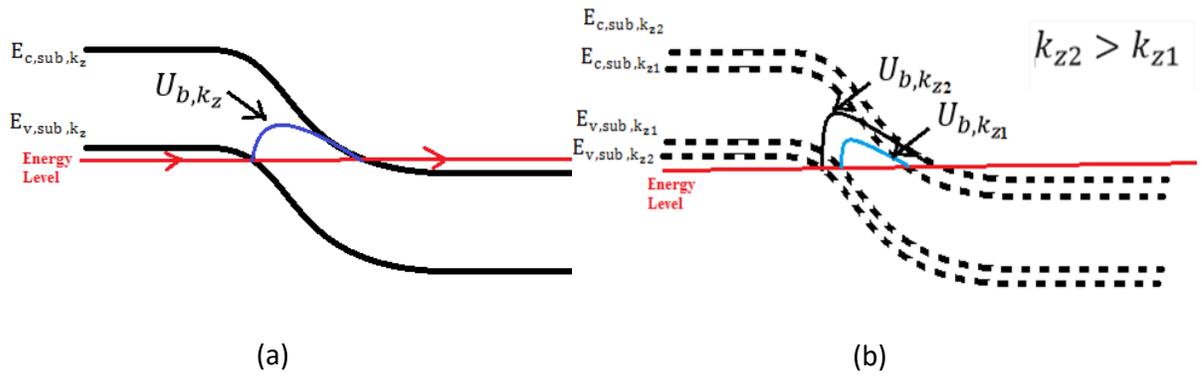

(a) (b)

**Figure 3.3: (a) The calculated barrier in the bandgap (b) Comparison of barrier for different values of $k_z$ i.e. different conduction and valence sub-band pairs.**

To understand why the effective bandgap increases with increasing value of $k_z$, let us take a look at Figure 3.4. We have shown the dispersion relationship for $E$ vs 2 components of $\vec{k}$- $k_x$ and $k_z$ as, obviously all 3 components cannot be shown with $E$ on a 3D plot. Two tunneling paths, one for $k_z = 0$ and another for $k_z \neq 0$ is shown. Notice that for $k_z \neq 0$ the electron has to travel a longer region in the bandgap. If we project the tunneling paths in the $E$ vs $k_x$ plane we see how the dispersion relationship clearly indicates an increase in effective bandgap for the sub-band corresponding to $k_z \neq 0$, and how the 2-band model approximates this increase.

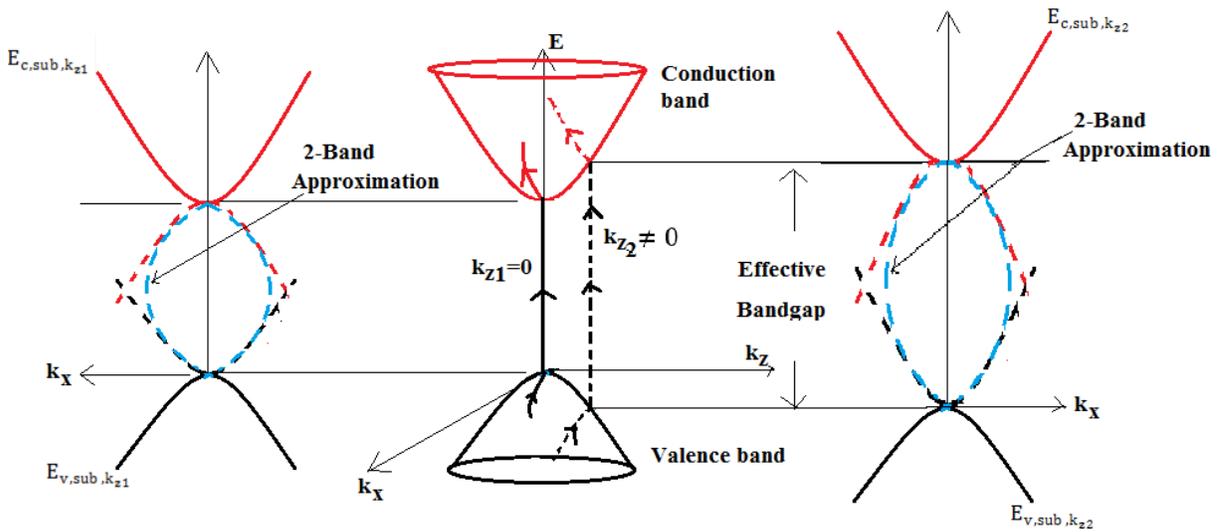

**Figure 3.4: The $E$ vs $k_x$ relation for two different values of $k_z$; showing the increase in effective bandgap for increased value of $k_z$**

Here, the tunneling effective mass $m_t^*(E)$ is dependent on the position of energy level of the electron in the bandgap, because it transitions from the value of hole effective mass in the valence band to electron effective mass in the conduction band. We will discuss how the effective mass varies in the bandgap in the next section.

## 3.4 Treatment of Tunneling Effective Mass

Like dispersion relationship in the bandgap, an exact treatment tunneling effective mass would require analysis of the band structure of the semiconductor material under consideration. However, we can make approximations for effective mass based on some logical assumptions.

Although various approximate schemes for evaluating tunneling effective mass can be used, several boundary conditions should be maintained. They are

$$m_t^*(E_v) = m_v^* \tag{3.46}$$

$$m_t^*(E_c) = m_c^* \tag{3.47}$$

Also, for a smooth transition from the effective mass approximation in the bands where effective masses are taken as constant with energy, we define two additional boundary conditions-

$$\left.\frac{\partial m_t^*}{\partial E}\right|_{E=E_v} = 0 \tag{3.48}$$

$$\left.\frac{\partial m_t^*}{\partial E}\right|_{E=E_v} = 0 \tag{3.49}$$

The simplest curve that satisfies 4 four boundary conditions is a cubic spline having a form of equation-

$$m_t^*(E) = a_0 + a_1 E + a_2 E^2 + a_3 E^3 \tag{3.50}$$

Here we have 4 constants which can be evaluated by applying the 4 boundary conditions to give a final form of the expression of effective mass,

$$m_t^*(E) = m_v^* + (m_c^* - m_v^*)\left[3\left(\frac{E-E_v}{E_c-E_v}\right)^2 - 2\left(\frac{E-E_v}{E_c-E_v}\right)^3\right] \tag{3.51}$$

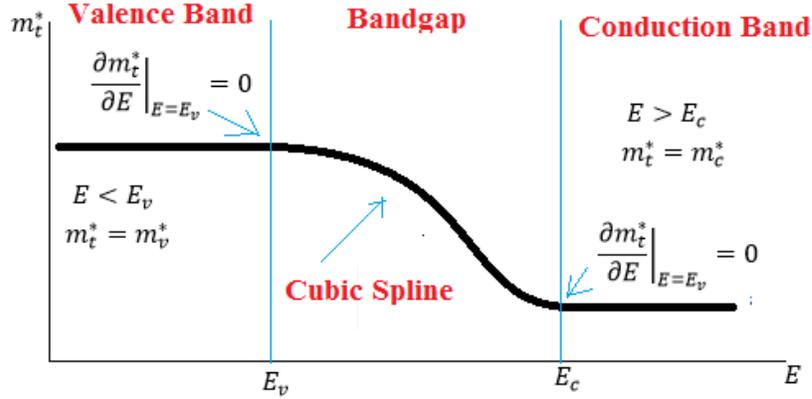

**Figure 3.5: Effective mass approximations in different energy ranges, the cubic spline connects hole and electron effective masses to give a smooth transition**

As shown in Figure 3.5, this allows a smooth transition of effective mass from valence to conduction band and also gets rid of any sharp discontinuity which would lead to inconsistent result in in the numerical solution.

For 2D potential profile the electron travels from a sub-band in valence band to a sub-band in conduction band. So, replacing $E_c$ and $E_v$ with $E_{c,sub,k_z}$ and $E_{v,sub,k_z}$ we get an appropriate expression for effective mass. The effective mass varies with space because it is dependent on the position of the energy level in the bandgap. Writing,

$$f(x,y) = \frac{E - E_{v,sub,k_z}(x,y)}{E_{c,sub,k_z}(x,y) - E_{v,sub,k_z}(x,y)} \qquad (3.52)$$

$$m_t^*(x,y) = m_v^* + (m_c^* - m_v^*)[3f(x,y)^2 - 2f(x,y)^3] \qquad (3.53)$$

One could of course approximate $m_t^*$ by a higher order polynomial of $E$ giving additional degrees of freedom for choosing the shape of the dispersion relationship in bandgap. However such modification and consequent increase in complexity may not be justified because of the approximate nature of the model.

## 3.5 Effective Potential and Mass in Different Regions

In order to extract parameters such as carrier concentration and current, we need to determine the potential profile seen by the electron during transport. The direct band to band tunneling process can be thought of as either transport of electron from valence to conduction band, or the transport of hole from conduction to valence band through a potential barrier. However we will consider electron as the tunneling carrier to explain the transport. We will also assume ballistic transport to simplify the transport phenomena.

While tunneling through the junction the electron first travels in the valence band, then enters the bandgap where it travels as an evanescent wave and emerges in the conduction band where it travels the rest of the distance. So, we define potential profile for these three different regions separately. We will consider a single band effective mass approximation in the conduction and valence band and the 2 band approximation in the bandgap to define the dispersion relationship.

For our 2D potential profile, as determined previously, in the valence band

$$\frac{\hbar^2}{2m_v^*}\nabla^2_{xy}\psi_{k_z}(x,y) + E_{v,\text{sub},k_z}(x,y)\psi_{k_z}(x,y) = E\psi_{k_z}(x,y) \quad (3.54)$$

The electron has a negative effective mass in the valence band. However, since for ballistic transport, energy $E$ remains the same during the transport time, the effect of negative effective mass can be explained in a different manner if we modify equation (3.54) as follows-

$$-\frac{\hbar^2}{2m_v^*}\nabla^2_{xy}\psi_{k_z}(x,y) + \overbrace{[2E - E_{v,\text{sub},k_z}(x,y)]}^{Effective\ potential,\ U_{v,kz}(x,y)}\psi_{k_z}(x,y) = E\psi_{k_z}(x,y) \quad (3.55)$$

Comparing this equation with the Schrodinger equation for a free electron, we see that for a particular energy level $E$, the effect of negative effective mass is accounted for if we change the effective potential to $[2E - E_{v,\text{sub},k_z}(x,y)]$. Then the electron appears to travel simply with a positive effective mass equal to $m_v^*$ i.e. the hole effective mass. So, we define the valence band effective potential as:

$$U_{v,k_z}(x,y) = 2E - E_{v,\text{sub},k_z}(x,y) \quad (3.56)$$

In the barrier the Schrodinger equation takes the form-

$$-\frac{\hbar^2}{2}\left[\frac{\partial}{\partial x}\left(\frac{1}{m_t^*(x,y)}\frac{\partial}{\partial x}\right) + \frac{\partial}{\partial y}\left(\frac{1}{m_t^*(x,y)}\frac{\partial}{\partial y}\right)\right]\psi_{k_z}(x,y) \\ + U_{b,k_z}(x,y)\psi_{k_z}(x,y) = E\psi_{k_z}(x,y) \quad (3.57)$$

Where $U_{b,k_z}(x,y)$ and $m_t^*(x,y)$ are determined for equations (3.45) and (3.53) respectively. The effective potential is simply $U_{b,kz}(x,y)$.

Finally in the conduction band, the effective potential is simply the sub-band energy $E_{c,sub,k_z}(x,y)$ and the effective mass is the electron effective mass $m_c^*$. That is-

$$U_{c,k_z}(x,y) = E_{c,sub,k_z}(x,y) \tag{3.58}$$

So, summarizing, the effective potential in different regions is given by

$$U_{k_z}(x,y) = \begin{cases} U_{v,kz}(x,y) \; ; where\ E \leq E_{v,sub,k_z}(x,y) \\ U_{b,kz}(x,y) \; ; where\ E_{v,sub,k_z}(x,y) < E < E_{c,sub,k_z}(x,y) \\ U_{c,kz}(x,y) \; ; where\ E \geq E_{c,sub,k_z}(x,y) \end{cases} \tag{3.59}$$

And the effective mass is given by

$$m_{k_z}^*(x,y) = \begin{cases} m_v^* \; ; where\ E \leq E_{v,sub,k_z}(x,y) \\ m_t^*(x,y) \; ; where\ E_{v,sub,k_z}(x,y) < E < E_{c,sub,k_z}(x,y) \\ m_c^* \; ; where\ E \geq E_{c,sub,k_z}(x,y) \end{cases} \tag{3.60}$$

Schrodinger equation for the electron over its whole transport length is given by

$$-\frac{\hbar^2}{2}\left[\frac{\partial}{\partial x}\left(\frac{1}{m_{k_z}^*(x,y)}\frac{\partial}{\partial x}\right) + \frac{\partial}{\partial y}\left(\frac{1}{m_{k_z}^*(x,y)}\frac{\partial}{\partial y}\right)\right]\psi_{k_z}(x,y) \\ + U_{k_z}(x,y)\psi_{k_z}(x,y) = E\psi_{k_z}(x,y) \tag{3.61}$$

This equation signifies the transition from a sub-band in the valence band to a sub-band in the conduction band both corresponding to a same value of $k_z$. The parameters $m_{k_z}^*(x,y)$ and $U_{k_z}(x,y)$ need to be calculated for each value of $k_z$ before this equation is solved. And the final value of any physical parameter, such as current or charge density is calculated by considering the contributions all possible values of $k_z$ i.e. all sub-bands.

As we have mentioned before, tunneling can be thought of as the transfer of electron or hole through the barrier. Figure 3.6 shows the effective potential for both type of carriers, but we will use only the one for electrons. Note however, how the effective potential for both carriers smoothly transitions from their respective sub-bands. This is achieved by combining the single band effective mass approximation with 2 band approximation. Also, note that we have so far used 1D plots (curves) for simple demonstration, actual potential profile is 2D and would require surface plots.

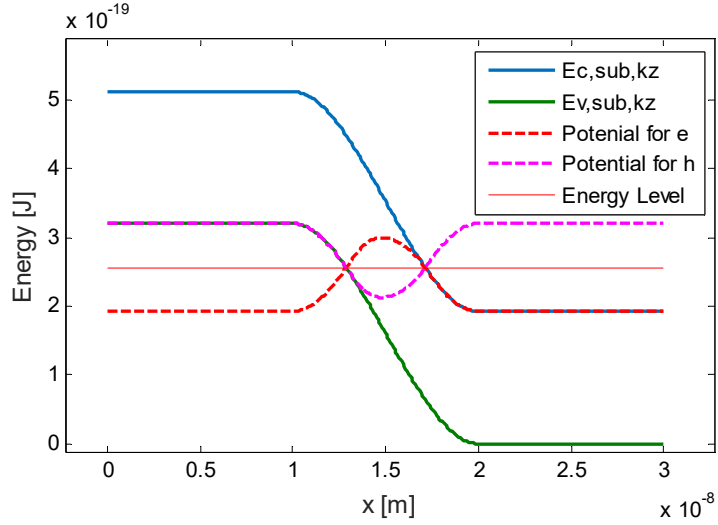

**Figure 3.6: Effective potential for both electron and hole in different regions**

## 3.6 Discretization and the Hamiltonian Matrix

To determine physical parameters like current and charge concentration involves solving equation (3.61) numerically for given boundary conditions. For this purpose we have to discretize the variables and form the Hamiltonian matrix.

We will show the discretization of the variables by Finite Difference Method (FDM) under central difference scheme [47]. For convenience we will now drop the subscript $k_z$.

Let's say, we want to divide the device into $N_x$ equal sections along the x axis and $N_y$ equal sections along the y axis. Let the spatial separation along $x$ and $y$ axis be $a_x$ and $a_y$ respectively. Each variable takes $(N_x + 1) \times (N_y + 1)$ discrete values at the mesh points. We will use the first subscript, say $i = 0, 1, \dots, N_x$ to denote the position along $x$ axis and the second subscript, say $j = 0, 1, \dots, N_y$ to denote position along y axis.

Now discretizing equation (3.61) under central difference scheme

$$-\frac{\hbar^2}{2}\left[\frac{1}{a_x^2}\left(\frac{\psi_{i+1,j} - \psi_{i,j}}{m^*_{i+\frac{1}{2},j}} - \frac{\psi_{i,j} - \psi_{i-1,j}}{m^*_{i-\frac{1}{2},j}}\right) + \frac{1}{a_y^2}\left(\frac{\psi_{i,j+1} - \psi_{i,j}}{m^*_{i,j+\frac{1}{2}}} - \frac{\psi_{i,i} - \psi_{i,j-1}}{m^*_{i,j-\frac{1}{2}}}\right)\right] \\ + U_{i,j}\psi_{i,j} = E\psi_{i,j} \quad (3.62)$$

We write the variables in column vector form-

$$\bar{\psi} = \{\psi_{i,j}\} = \begin{bmatrix} \psi_{0,0} \\ \psi_{0,1} \\ \psi_{0,2} \\ \vdots \\ \psi_{0,N_y} \\ \psi_{1,0} \\ \psi_{1,1} \\ \vdots \\ \vdots \\ \psi_{N_x,N_y} \end{bmatrix} \quad ; \quad \bar{U} = \{U_{i,j}\} = \begin{bmatrix} U_{0,0} \\ U_{0,1} \\ U_{0,2} \\ \vdots \\ U_{0,N_y} \\ U_{1,0} \\ U_{1,1} \\ \vdots \\ \vdots \\ U_{N_x,N_y} \end{bmatrix}$$

The 2D device Hamiltonian matrix in the absence of external potential is denoted by $H_0$. It is a sparse matrix of size $= (N_y + 1)(N_x + 1) \times (N_y + 1)(N_x + 1)$ with 5 non-zero diagonals. The form of the matrix is shown in Figure 3.7. The different elements are determined by the values,

$$\varepsilon_{i,j} = t_{x,i,j} + t_{x,i+1,j} + t_{y,i,j} + t_{x,i,j+1} \tag{3.63}$$

$$t_{x,i,j} = \frac{\hbar^2}{2m^*_{i-\frac{1}{2},j} a_x^2} \tag{3.64}$$

$$t_{y,i,j} = \frac{\hbar^2}{2m^*_{i,j-\frac{1}{2}} a_y^2} \tag{3.65}$$

In particular we define-

$$t_{x,0,j} = t_{x,1,j} \tag{3.66}$$

$$t_{x,N_x+1,j} = t_{x,N_x,j} \tag{3.67}$$

$$t_{y,i,0} = t_{y,i,1} \tag{3.68}$$

$$t_{y,i,N_y+1} = t_{y,i,N_y} \tag{3.69}$$

$$H_0 = \begin{bmatrix} \varepsilon_{0,0} & -t_{y,0,1} & & & -t_{x,1,0} & & & & & & & & \\ -t_{y,0,1} & \varepsilon_{0,1} & -t_{y,0,2} & & & -t_{x,1,1} & & & & & & & \\ & \ddots & \ddots & \ddots & & & \ddots & & & & & & \\ & & -t_{y,0,N_y} & \varepsilon_{0,N_y} & & & & -t_{x,1,N_y} & & & & & \\ -t_{x,1,0} & & & 0 & \varepsilon_{1,0} & -t_{y,1,1} & & & -t_{x,2,0} & & & & \\ & -t_{x,1,1} & & & -t_{y,1,1} & \varepsilon_{1,1} & -t_{y,1,2} & & & -t_{x,2,1} & & & \\ & & \ddots & & & \ddots & \ddots & \ddots & & & \ddots & & \\ & & & -t_{x,1,N_y} & & & & & & & & -t_{N_x,N_y-1} & \varepsilon_{N_x,N_y-1} & -t_{y,N_x,N_y} \\ & & & & & & & & & & & -t_{y,N_x,N_y} & \varepsilon_{N_x,N_y} \end{bmatrix}$$

**Figure 3.7:** The 2D device Hamiltonian matrix in the absence of external potential, $H_0$

In presence of external potential, the Hamiltonian matrix takes the form

$$H = H_0 + diag(\bar{U}) \tag{3.70}$$

That is, the elements of vector $\bar{U}$ are added to the diagonal elements of the Hamiltonian matrix.

So, the equation (3.61) when discretized, takes the form:

$$H\bar{\psi} = E\bar{\psi} \tag{3.71}$$

However, we must consider proper boundary conditions to model the device. Open boundary condition is applied at the source and drain to take into effect the transmission of electrons. This is where NEGF formalism comes to our aid.

## 3.7 The Choice of a Solution Method: 2D NEGF

Now that we have discretized the Schrodinger's equation, we need to solve it numerically. However to do so, we need to apply appropriate boundary conditions. In our case, the boundary condition is closed everywhere except at the leads i.e. at source and drain where the boundary is open which means electron wave function can flow in or out. So, we choose NEGF formalism which gives a sound mathematical solution for transmitting boundary conditions [48].

The direct band to band tunneling is an elastic process. So it can modeled very easily with NEGF formalism, which integrates this process with ballistic transport. Unlike WKB approximation, the 2D NEGF formalism can give us the exact transmission coefficient for 2D potential without any assumption of tunneling direction. In case of quasi-ballistic transport dephasing mechanisms can be added to model scattering [49]. Quantum mechanical effects like confinement and intra-band tunneling are easily accounted for because we are directly solving Schrodinger's equation.

Before proceeding, we clarify why we did not use a more computationally efficient method like Uncoupled Mode Space (UMS). One of the simplifying assumptions of UMS with 1D NEGF approach is that the confining potential profile is fairly invariant along the transport direction [50]. This is only true for a fairly 1D potential profile (like HEMT or Nanowire), but this beats our purpose for developing a 2D Model. We could of course use Coupled Mode Space (CMS) approach to remedy this, but it is almost equal in complexity as 2D NEGF. Although at first look, 2D NEGF formalism may seem computationally less efficient than CMS because larger matrices are involved in computation, they are also sparser and it can be optimized for faster simulation speed when there is lateral confinement by selecting a few sub-bands [51] which is explained in Appendix B.

When we are using the NEGF method, we are basically treating the device as a waveguide (Figure 3.8), so that the wave nature of electron is accounted for. This is different from treating the device as a combination of parallel tunneling current paths as done in the semi-classical approach. The leads (source and drain) are considered incoherent sources.

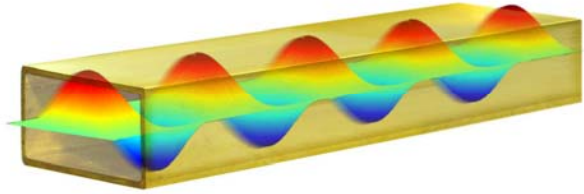

**Figure 3.8: The waveguide model**

Just like the waveguide has a cutoff frequency for each mode, below which waves cannot pass, electrons traveling through a highly confined region have minimum energy levels below which they cannot pass, and this signifies the effect of confinement on electric current.

## 3.8 The Non-Equilibrium Green's Function Formalism

The Non-Equilibrium Green's Function [48] is a mathematical formalism for solving the Schrodinger equation for open boundary condition. In this method, the Hamiltonian matrix *H* models the device, just like the closed boundary Schrodinger's equation. But open boundary condition means the leads, in our case the source and the drain must be considered semi-infinite. The wave function can flow in and out through the leads.

We cannot make the Hamiltonian Matrix semi-infinite. But it can be collapsed into a finite matrix by absorbing the self-energy into the first element [52]. This is taken care of by the Self-energy Matrices $\Sigma_1$ and $\Sigma_2$.

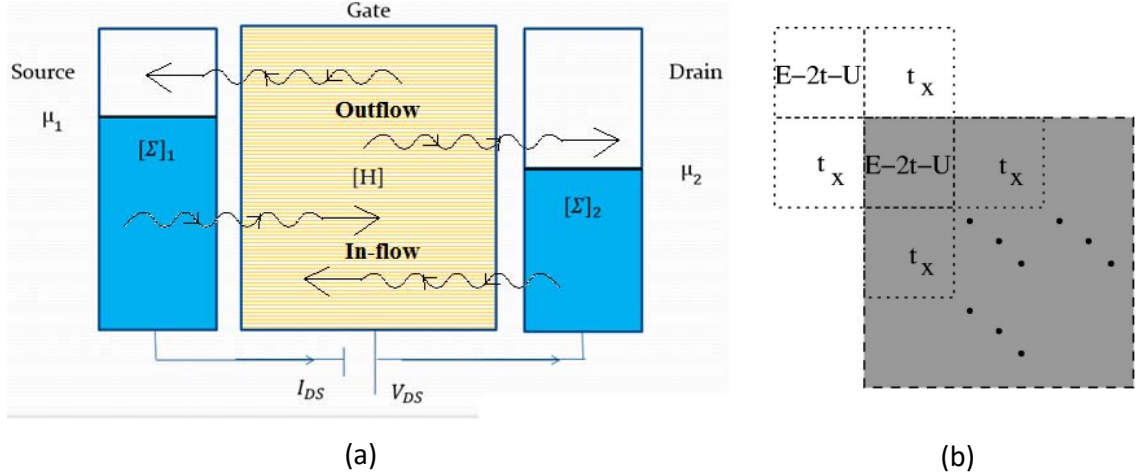

**Figure 3.9: (a) How NEGF formalism models different parts of device and transport phenomena; (b) How the semi-infinite H matrix can be collapsed to get the self-energy term**

The inflow can be calculated from equilibrium condition, i.e. when inflow and outflow are equal. Difference in source and drain fermi levels drive the current. Self-energy matrices are also used to model scattering [53] i.e. dephasing mechanisms which are equivalent to removing and reinserting the electrons in the device. The corresponding self-energy matrix is denoted by $\Sigma_S$.

## 3.9 The Basic Equations of NEGF Formalism

In this section we will primarily discuss the equations for a particular value of $k_z$ and drop the subscript for simplicity. However we will add the $k_z$ subscript to signify contributions from different values of $k_z$ in the end results.

The Green's function, also called the retarded Green's function [51] is given by

$$G(E) = G^R(E) = \left[ ES - H(E) - \sum_\alpha \Sigma_\alpha(E) \right]^{-1} \qquad (3.72)$$

Where, in our case $S$ is just the identity matrix. (We will keep using this notation for Identity matrix to avoid conflict with electric current $I$.) $\Sigma_\alpha(E)$ are the self-energy matrices. Ignoring scattering there are only two self-energy matrices need to be considered, $\Sigma_1$ and $\Sigma_2$ for source and drain respectively. So,

$$G(E) = [ES - H(E) - \Sigma_1(E) - \Sigma_2(E)]^{-1} \tag{3.73}$$

Method of calculation of the self-energy matrices is explained in Appendix A.

The advanced Green's function is the conjugate transpose of retarded Green's function

$$G^A(E) = G^\dagger(E) \tag{3.74}$$

The spectral density functions are given by,

$$A_1(E) = G(E)\Gamma_1(E)G^\dagger(E) \tag{3.75}$$
$$and, A_1(E) = G(E)\Gamma_1(E)G^\dagger(E) \tag{3.76}$$

Where the broadening functions $\Gamma_1$ and $\Gamma_2$ are given by,

$$\Gamma_1(E) = j[\Sigma_1(E) - \Sigma_1^\dagger(E)] \tag{3.77}$$
$$\Gamma_2(E) = j[\Sigma_2(E) - \Sigma_2^\dagger(E)] \tag{3.78}$$

The non-equilibrium density matrix

$$P = \frac{1}{2\pi} \int_{-\infty}^{+\infty} [f(\mu_1, E)A_1(E) + f(\mu_2, E)A_2(E)]dE \tag{3.79}$$

Where, $\mu_1$ and $\mu_2$ are source and drain Fermi levels respectively. $f(\mu, E)$ is the Fermi-Dirac distribution function given by

$$f(\mu, E) = \frac{1}{1 + \exp\left(\frac{E - \mu}{kT}\right)} \tag{3.80}$$

We don't need all the elements of the density matrix P. To simplify we define two new variables $D_1$ and $D_2$ which signify Local Density of States (LDOS) due to source and drain respectively. They correspond to the diagonal elements of their respective spectral density matrices. A spin degeneracy factor of 2 is added and the column vector is normalized with respect to spatial coordinates.

$$D_1[p] = \frac{1}{\pi \times a_x \times a_y} A_1[p, p] \tag{3.81}$$

$$D_2[p] = \frac{1}{\pi \times a_x \times a_y} A_2[p, p] \tag{3.82}$$

After determination of the LDOS, we may proceed to calculate the carrier concentration. The 2D electron concentration is given by

$$n^{2D} = \int_{k_z=0}^{\infty} \int_{E=-\infty}^{+\infty} [D_{1,k_z}(E)f(\mu_1,E) + D_{2,k_z}(E)f(\mu_2,E)]\rho(k_z)dE \quad (3.83)$$

$$= \int_{k_z=0}^{\infty} \int_{E=-\infty}^{\infty} [D_{1,k_z}(E)f(\mu_1,E) + D_{2,k_z}(E)f(\mu_2,E)]\left(\frac{L_z}{\pi}\right)dE\, dk_z \quad (3.84)$$

Here, we have re-introduced the subscript '$k_z$' to denote the contribution from sub-bands corresponding to particular value of $k_z$ and inserted the value of $\rho(k_z)$ from equation (3.15).

Now, the 3D electron density is

$$n = \frac{n^{2D}}{L_z} = \frac{1}{\pi}\int_{k_z=0}^{\infty}\int_{E=-\infty}^{\infty}[D_{1,k_z}(E)f(\mu_1,E) + D_{2,k_z}(E)f(\mu_2,E)]dE\, dk_z \quad (3.85)$$

Similarly, the hole density is given by

$$p = \frac{1}{\pi}\int_{k_z=0}^{\infty}\int_{E=-\infty}^{\infty}[D_{1,k_z}(E)\{1-f(\mu_1,E)\} + D_{2,k_z}(E)\{1-f(\mu_2,E)\}]dE\, dk_z \quad (3.86)$$

The current is given by

$$I = \frac{q}{\pi\hbar}\int_{k_z=0}^{k_{z,max}}\int_{E=E_{min}}^{E_{max}} T_{k_z}(E)[f(\mu_1,E)-f(\mu_2,E)]\left(\frac{L_z}{\pi}\right)dE\, dk_z \quad (3.87)$$

Current per unit width of the device is given by

$$\frac{I}{L_z} = \frac{q}{\pi^2\hbar}\int_{k_z=0}^{k_{z,max}}\int_{E=E_{min}}^{E_{max}} T_{k_z}(E)[f(\mu_1,E)-f(\mu_2,E)]dE\, dk_z \quad (3.88)$$

Where, Transmission coefficient,

$$T_{k_z}(E) = trace[\Gamma_{1,k_z}(E)G_{k_z}(E)\Gamma_{1,k_z}(E)G_{k_z}^{\dagger}(E)] \quad (3.89)$$

Note that the NEGF formalism allows us to calculate the transmission coefficient for the whole device without any prior assumption of tunneling path.

Also note that to get the current per unit width, we need to integrate over both energy and momentum, $k_z$. But as we will see in the next chapter we only need to do this once and over a finite limit at the end of the simulation, so time is not the limiting factor. But, notice that both of the expressions for carrier concentration contain double integrals which extend to infinity. This calculation can be very time consuming, especially in a self-consistent loop. So we will simplify these expressions in section 3.11.

The evaluation of the Green's function matrix $G$ is a time consuming process because it involves matrix inversion. This is especially problematic in case of 2D NEGF. But calculations can be made much faster by partial evaluation of the Green's function [51] as explained in Appendix A. The equations given in this section will take a slightly different form if we perform this modification.

## 3.10 Limits of Integration

Range of energy, over which tunneling occurs

$$min\left(E_{c,sub,k_z}(x,y)\right) < E < max\left(E_{v,sub,k_z}(x,y)\right) \tag{3.90}$$

So,

$$E_{min} = min\left(E_{c,sub,k_z}(x,y)\right) \tag{3.91}$$

$$E_{max} = max\left(E_{v,sub,k_z}(x,y)\right) \tag{3.92}$$

As for range of $k_z$ over which tunneling may occur

$$E_{min} + \frac{\hbar^2 k_{z,max}^2}{2m_c^*} = E_{max} - \frac{\hbar^2 k_{z,max}^2}{2m_c^*} \tag{3.93}$$

$$\Rightarrow k_{z,max} = \frac{\sqrt{2m_r^*(E_{max} - E_{min})}}{\hbar} \tag{3.94}$$

Where, the reduced effective mass is

$$m_r^* = \frac{m_c^* m_v^*}{m_c^* + m_v^*} \tag{3.95}$$

## 3.11 Simplified Approximation of Carrier Concentration

Assuming that the source is p type and the drain is n type and a significant tunneling barrier between conduction and valence band it is a good approximation to assume that only the source fermi level contributes to holes and the drain fermi level contributes to electrons.

So we get simplified expressions for electron and hole concentrations

$$n = \frac{1}{\pi} \int_{k_z=0}^{\infty} \int_{E=-\infty}^{+\infty} D_{2,k_z}(E) f(\mu_2, E) dE\, dk_z \qquad (3.96)$$

$$p = \frac{1}{\pi} \int_{k_z=0}^{\infty} \int_{E=-\infty}^{\infty} D_{1,k_z}(E) \{1 - f(\mu_1, E)\} dE\, dk_z \qquad (3.97)$$

Now, the dependence of $D_1$ and $D_2$ on $k_z$ makes the computational process very lengthy when the double integral needs to be carried out numerically. This significantly slows down the self-consistent loop (which is to be discussed later). The speed can be increased if we make a simplifying assumption which is discussed next.

First, to simplify electron concentration let us make a change of variable. Let us take,

$$E = E_\parallel + \frac{\hbar^2 k_z^2}{2m_c^*} \Rightarrow E_\parallel = E - \frac{\hbar^2 k_z^2}{2m_c^*} \qquad (3.98)$$

$$Then,\, dE = dE_\parallel \qquad (3.99)$$

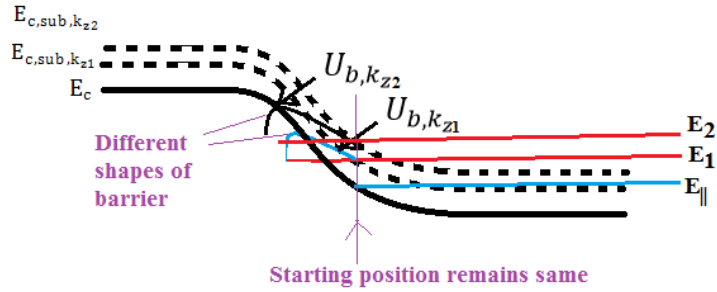

**Figure 3.10: Dependence of effective potential on $k_z$ for a given value of $E_\parallel$**

Where $E_\parallel$ is the energy excluding the energy in the z direction. Now let's observe how the effective potential changes for a particular value of $E_\parallel$ but different values of $k_z$ (Figure 3.10). The height and shape of the barrier ($U_{b,k_z}$) changes with $k_z$ but its starting position remain the same. The shape of the barrier is important for current calculation but not for carrier calculation because the carrier concentration decays very rapidly inside the barrier. Only where the barrier starts is important.

So, $D_{2,k_z}(E)$ can be assumed to be almost equal for a particular value of $E_\parallel$ but different values of $k_z$. Which allows us to write,

$$D_{2,k_z}(E) = D_2(E_\parallel) = D_{2,k_z}(E_\parallel)\big|_{k_z=\infty} \qquad (3.100)$$

Where we have arbitrarily chosen $k_z = \infty$, which means absence of BTBT. So, we have effectively assumed that the BTBT doesn't change the equilibrium carrier concentration very much. Now, electron concentration becomes,

$$n = \frac{1}{\pi} \int_{k_z=0}^{\infty} \int_{E=E_{c,min}}^{\infty} D_2(E_\parallel) f\left(\mu_2, E_\parallel + \frac{\hbar^2 k_z^2}{2m_c^*}\right) dE_\parallel \, dk_z$$

$$= \int_{E=E_{c,min}}^{\infty} D_2(E_\parallel) \left[\frac{1}{\pi} \int_{k_z=0}^{\infty} f\left(\mu_2, E_\parallel + \frac{\hbar^2 k_z^2}{2m_c^*}\right) dk_z\right] dE_\parallel$$

$$So, n = \int_{E=E_{c,min}}^{\infty} D_2(E_\parallel) f_c(\mu_2, E_\parallel) dE_\parallel \tag{3.101}$$

Where we define,

$$f_c(\mu_2, E_\parallel) = \frac{1}{\pi} \int_{k_z=0}^{\infty} f\left(\mu_2, E_\parallel + \frac{\hbar^2 k_z^2}{2m_c^*}\right) dk_z$$

$$= \sqrt{\frac{m_c^* kT}{2\pi \hbar^2}} \mathcal{F}_{-\frac{1}{2}}\left(\frac{\mu_2 - E_\parallel}{kT}\right) \tag{3.102}$$

Equation (3.101) requires only one integration and thus computationally much faster than the double integration required for equation (3.96). We need the Fermi-Dirac integral to evaluate this expression, which is given by

$$\mathcal{F}_k(x) = \frac{1}{\Gamma(k+1)} \int_0^\infty \frac{t^k dt}{\exp(t-x)+1}, \quad k > -1 \tag{3.103}$$

The values for Fermi-Dirac integral is taken from tabulated values, thus do not consume much time to evaluate.

We can make similar simplification for hole concentration taking

$$E_\parallel = E + \frac{\hbar^2 k_z^2}{2m_v^*} \tag{3.104}$$

The hole concentration is given by

$$p = \int_{E=-\infty}^{E_{v,max}} D_1(E_\parallel) \left[\frac{1}{\pi} \int_{k_z=0}^{\infty} \left\{1 - f\left(\mu_1, E_\parallel - \frac{\hbar^2 k_z^2}{2m_v^*}\right)\right\} dk_z\right] dE_\parallel$$

$$\text{So}, p = \int_{E=-\infty}^{E_{v,max}} D_2(E_\parallel) f_v(\mu_1, E_\parallel) dE_\parallel \tag{3.105}$$

Where,

$$\begin{aligned} f_v(\mu_1, E_\parallel) &= \frac{1}{\pi} \int_{k_z=0}^{\infty} \left\{ 1 - f\left(\mu_1, E_\parallel - \frac{\hbar^2 k_z^2}{2m_v^*}\right) \right\} dk_z \\ &= \sqrt{\frac{m_v^* kT}{2\pi \hbar^2}} \mathcal{F}_{-\frac{1}{2}}\left(\frac{E_\parallel - \mu_1}{kT}\right) \end{aligned} \tag{3.106}$$

## 3.12 Chapter Summary

In this chapter we have shown the step by step mathematical deduction that lead to the development of our model. We have also introduced the 2D NEGF formalism for numerical solution and simplified the results obtained from this formalism. In the next chapter we will discuss, how the results obtained in this chapter may be used in simulation.

# Chapter 4
# Simulation Methodology

In this chapter we will discuss how our numerical model may be implemented in simulation of TFETs. After a generalized discussion we will demonstrate a practical implementation of our model for simulation of a double gate TFET. The results of the performed simulation will be discussed in the next chapter.

## 4.1 Summary of Important Equations

First we summarize the equations required for simulation from the previous chapter.

Sub-band energies:

$$E_{v,sub,k_z}(x,y) = E_v(x,y) - \frac{\hbar^2 k_z^2}{2m_v^*} \tag{4.1}$$

$$E_{c,sub,k_z}(x,y) = E_c(x,y) + \frac{\hbar^2 k_z^2}{2m_c^*} \tag{4.2}$$

Tunneling barrier:

$$U_{b,k_z}(x,y) = E - \frac{\hbar^2 k_{xy}^2(x,y)}{2m_t^*(E)} \tag{4.3}$$

Tunneling effective mass:

$$f(x,y) = \frac{E - E_{v,sub,k_z}(x,y)}{E_{c,sub,k_z}(x,y) - E_{v,sub,k_z}(x,y)} \tag{4.4}$$

$$m_t^*(x,y) = m_v^* + (m_c^* - m_v^*)[3f(x,y)^2 - 2f(x,y)^3] \tag{4.5}$$

Schrodinger equation for particular value of $k_z$:

$$-\frac{\hbar^2}{2}\left[\frac{\partial}{\partial x}\left(\frac{1}{m_{k_z}^*(x,y)}\frac{\partial}{\partial x}\right) + \frac{\partial}{\partial y}\left(\frac{1}{m_{k_z}^*(x,y)}\frac{\partial}{\partial y}\right)\right]\psi_{k_z}(x,y) \\ + U_{k_z}(x,y)\psi_{k_z}(x,y) = E\psi_{k_z}(x,y) \tag{4.6}$$

The Hamiltonian Matrix:

$$H = H_0 + diag(\overline{U}) \tag{4.7}$$

The NEGF Formalism:

- ❖ The (Retarded) Green Function:

$$G(E) = [ES - H(E) - \Sigma_1(E) - \Sigma_2(E)]^{-1} \qquad (4.8)$$

- ❖ The Spectral Density Functions:

$$A_1(E) = G(E)\Gamma_1(E)G^\dagger(E) \qquad (4.9)$$

$$and, \ A_1(E) = G(E)\Gamma_1(E)G^\dagger(E) \qquad (4.10)$$

- ❖ Broadening functions:

$$\Gamma_1(E) = j[\Sigma_1(E) - \Sigma_1^\dagger(E)] \qquad (4.11)$$

$$\Gamma_2(E) = j[\Sigma_2(E) - \Sigma_2^\dagger(E)] \qquad (4.12)$$

- ❖ The LDOS:

$$D_1[p] = \frac{1}{\pi \times a_x \times a_y} A_1[p,p] \qquad (4.13)$$

$$D_2[p] = \frac{1}{\pi \times a_x \times a_y} A_2[p,p] \qquad (4.14)$$

- ❖ The carrier concentrations:

$$n = \int_{E=E_{c,min}}^{\infty} D_2(E_\parallel) f_c(\mu_2, E_\parallel) dE_\parallel \qquad (4.15)$$

$$p = \int_{E=-\infty}^{E_{v,max}} D_2(E_\parallel) f_v(\mu_1, E_\parallel) dE_\parallel \qquad (4.16)$$

- ❖ The current per unit width:

$$\frac{I}{L_z} = \frac{q}{\pi^2 \hbar} \int_{k_z=0}^{k_{z,max}} \int_{E=E_{min}}^{E_{max}} T_{k_z}(E)[f(\mu_1, E) - f(\mu_2, E)] dE \, dk_z \qquad (4.17)$$

- ❖ The transmission coefficient:

$$T_{k_z}(E) = trace[\Gamma_{1,k_z}(E) G_{k_z}(E) \Gamma_{1,k_z}(E) G_{k_z}^\dagger(E)] \qquad (4.18)$$

## 4.2 The Self-consistent Loop

Before physical parameters like current or carrier concentration can be calculated we must establish consistency between Schrodinger's equation and the Poisson's equation. The Poisson's equation is given by

$$-\nabla.[\epsilon \nabla V(x,y)] = \rho(x,y) \qquad (4.19)$$

Here, $V(x,y)$ is the electric potential and $\rho(x,y)$ is the charge density. The charge density is calculated from the carrier concentrations calculated by 2D NEGF formalism as-

$$\rho(x,y) = e(p - n + N_D^+ - N_A^-) \qquad (4.20)$$

Here, $N_D^+$ and $N_A^-$ are charge concentration due to ionized donors and acceptors respectively and $n$ and $p$ comes for equations (4.15) and (4.16) respectively.

Applying proper boundary conditions and solving numerically, Poisson's equation yields the potential profile $V(x,y)$, which is used to calculate the bands $E_c(x,y)$ and $E_v(x,y)$ which in turn is again used in 2D NEGF formalism. This loop is continued until consistency is achieved. Either carrier relaxation or band relaxation method or a combination of the two can be used for convergence.

For carrier relaxation-

$$n_{new} = (1-\alpha)n_{calculated} + \alpha n_{old} \qquad (4.21)$$
$$p_{new} = (1-\alpha)p_{calculated} + \alpha p_{old} \qquad (4.22)$$

For band relaxation-

$$E_{c,new} = (1-\alpha)E_{c,calculated} + \alpha E_{c,old} \qquad (4.23)$$
$$E_{c,new} = (1-\alpha)E_{v,calculated} + \alpha E_{v,old} \qquad (4.24)$$

Here $\alpha$ is a forgetting factor which is adjusted for faster convergence and to prevent divergenc.

After consistency is achieved current and other parameters can be calculated.

## 4.3 Simulation Workflow

The flowchart for simulation is shown in Figure 4.1. The simulation workflow goes through the following steps:

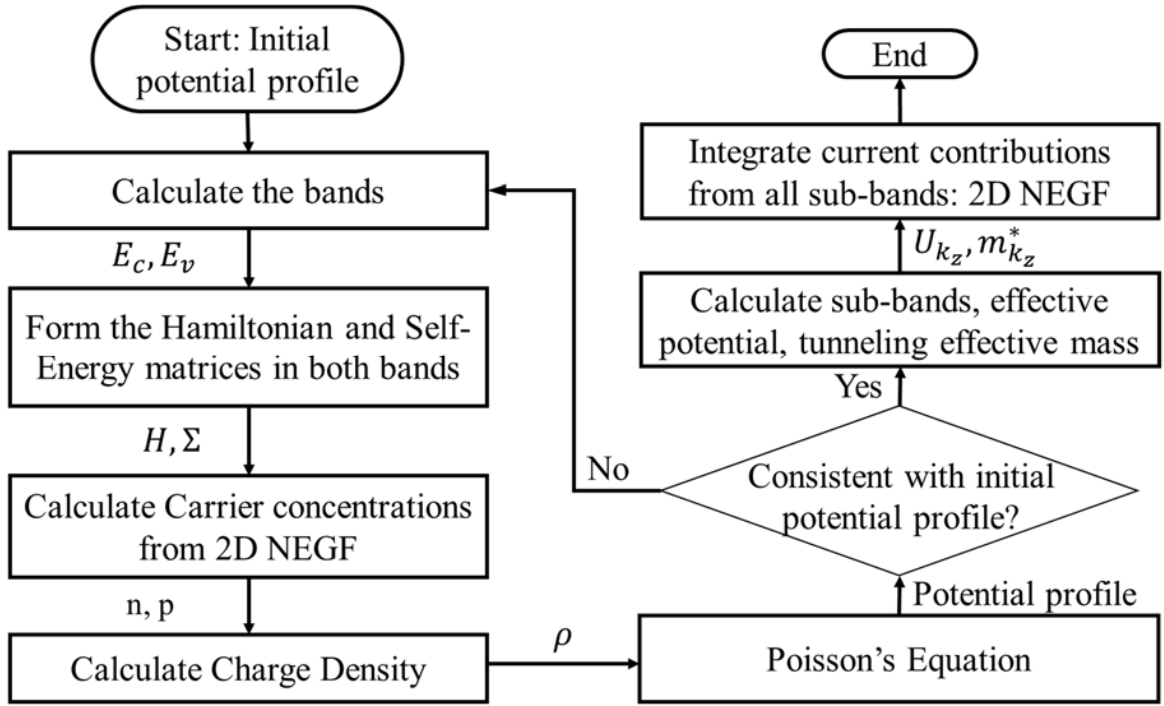

**Figure 4.1: Simulation Flowchart**

Step 1: We start with an initial potential profile, $V(x,y)$. Instead of starting with a random profile we can start with an intelligent guess which will reduce the time taken for convergence.

Step 2: From the potential profile we calculate the band profile $E_c(x,y)$ and $E_v(x,y)$.

Step 3: We form the Hamiltonian matrix $H$ and self-energy matrices $\Sigma_1$ and $\Sigma_2$ in both conduction and valence bands.

Step 4: We use the 2D NEGF formalism to calculate the electron and hole concentrations $n$ and $p$ respectively.

Step 5: From the carrier concentration and doping profile we calculate the charge density $\rho(x,y)$.

Step 6: The charge density is used in Poisson's equation to calculate a new potential profile.

Step 7: If the new potential profile is within a tolerance limit of the old potential profile, we have reached convergence and we may proceed to the next step. But if the potential profile is not consistent with the initial profile, we go back to step 2 and use the new potential profile to calculate the bands. This loop is iterated until convergence i.e. self-consistency is achieved.

Step 8: Once convergence is reached we calculate the sub-band profiles $E_{c,sub,k_z}$ and $E_{v,sub,k_z}$, effective potential $U_{k_z}$ and $m^*_{k_z}$ for different values of $k_z$ within the BTBT range.

Step 9: Finally we use the 2D NEGF formalism to integrate the current contributions from all sub-bands i.e. all values of $k_z$ to get the final value of current per unit length.

## 4.4 Dealing with Potential Pockets

If the band profiles contain potential pockets i.e. confined regions in the transport direction in the final or an intermediate step in the self-consistent loop, the ballistic NEGF formalism predicts this pocket to be empty of carriers [41], because it does not consider any scattering or generation recombination process (Figure 4.2). This may lead to problem in achieving self-consistency by making the charge distribution vastly different in two successive steps, due to formation of potential pockets. Even if consistency is achieved, the predicted charge concentration may be different from real values, because in reality the potential pockets should eventually fill up because of scattering or thermal generation process.

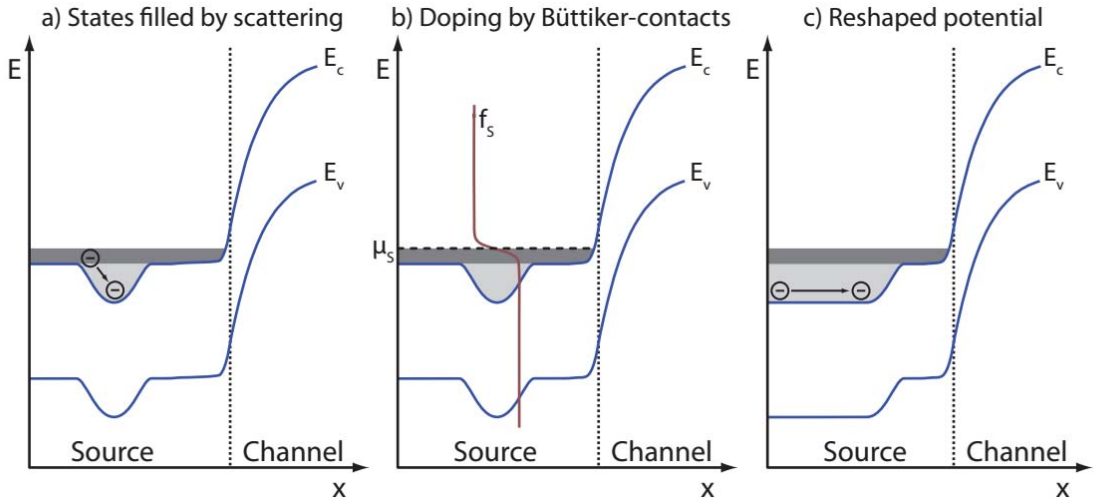

**Figure 4.2: Dealing with potential pockets**

A proper treatment of this problem exists within the NEGF formalism, where Buttiker contacts are placed within the potential well to model scattering [53, 54]. But this approach would consume significantly more time than the ballistic model. So, a faster and simpler solution is to artificially remove the potential pocket so that carriers may flow into this region [41]. Another solution by closed boundary Schrodinger equation for high scattering is presented in the next section

## 4.5 Simplification of Simulation Process

To simplify the simulation process, different tactics can be adopted in different cases:

- For faster convergence of the self-consistent loop, the initial potential profile can be taken from a semi-classical drift-diffusion simulation. The semi-classical simulation can be very simplistic in nature and may ignore BTBT as we are using it just for an initial guess. The task then reduces to fine tuning that initial profile found from semi-classical simulation which is much easier than starting from scratch.
- If the device is reasonably thick, say thicker than 15 nm, and the potential profile shows no signs of confined regions, we may overlook lateral confinement. Then we can take the band profile from a semi-classical simulation, skip the self-consistent loop and proceed to calculate current by 2D NEGF. This mixed process gives a better prediction of current than semi-classical methods because 2D NEGF does not require any assumption of tunneling path. However we must be cautious because this method has the risk of ignoring any carrier confinement due to potential distribution.
- To add the effect of carrier confinement but for faster convergence and quicker simulation, the open boundaries can be replaced with closed boundaries in the self-consistent loop. The problem then reduces to a conventional Eigenvalue problem and NEGF is not required to calculate carriers, which makes the self-consistent loop much faster. This method may be more appropriate for a device where scattering is dominant and ballistic transport only applies within the tunneling distance.
  However to account for the open-boundary condition and the semi-infinite nature of the leads, we may extend the device Hamiltonian a little longer beyond the leads as shown in Figure 4.3, and then perform the carrier calculations.

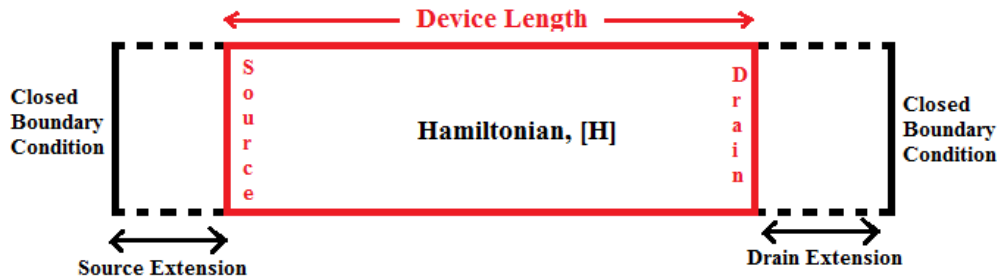

**Figure 4.3: Imitating semi-infinite leads by extending the Hamiltonian H beyond the leads**

## 4.6 A Practical Implementation

To experiment with the implications of our model and it's performance level in TFET simulation with respect to other models we will do a simulation for a practical TFET structure given in [40]. We will experiment on the TFET structure of shown in Figure 4.4. It is a double gate Sillicon p+-n-n+ TFET. The highly doped p+ region ($10^{20}$ atoms/cm$^3$), the lightly n-doped intermediate region ($10^{17}$ atoms/cm$^3$), and the highly doped n+ region ($10^{20}$ atoms/cm$^3$) act as the source, the channel, and the drain, respectively. Metal gates with work function of 4.5 $eV$ are placed over and below the channel with a length of 20 nm. A 1-nm-thick SiO$_2$ layer is used as gate oxide at both gates.

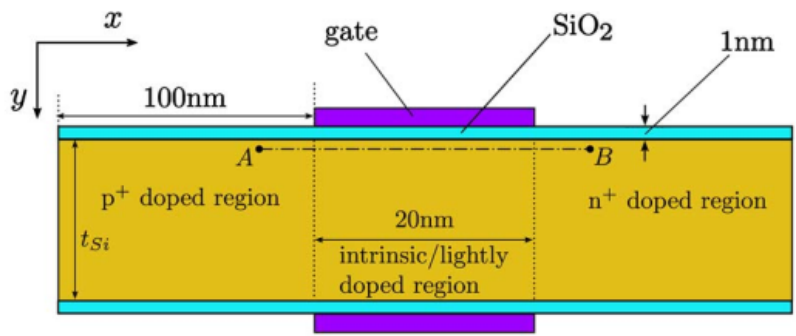

**Figure 4.4: The TFET device structure to be experimented on**

Simulation results for semi-classical and quasi-quantum models of this structure is give in [40]. We will simulate the device for different drain, gate voltages and thickness and compare the result with that of semi-classical and quasi-quantum models.

## 4.7 Implementation of the Simulator

To implement our model in simulation we used two software packages, COMSOL Multiphysics 5.2 and Matlab 2015b. We used the graphical user interface of COMSOL Multiphysics to design the TFET structure, parameters and boundary conditions. We used the PDE solvers of COMSOL Multiphysics perform the semi-classical simulation and also to solve the Poisson's equation in the self-consistent loop. We used Matlab to implement the 2D NEGF formalism and carry out all other mathematical operations.

Figure 4.5 shows the flowchart followed by the simulator and the operations performed in different stages.

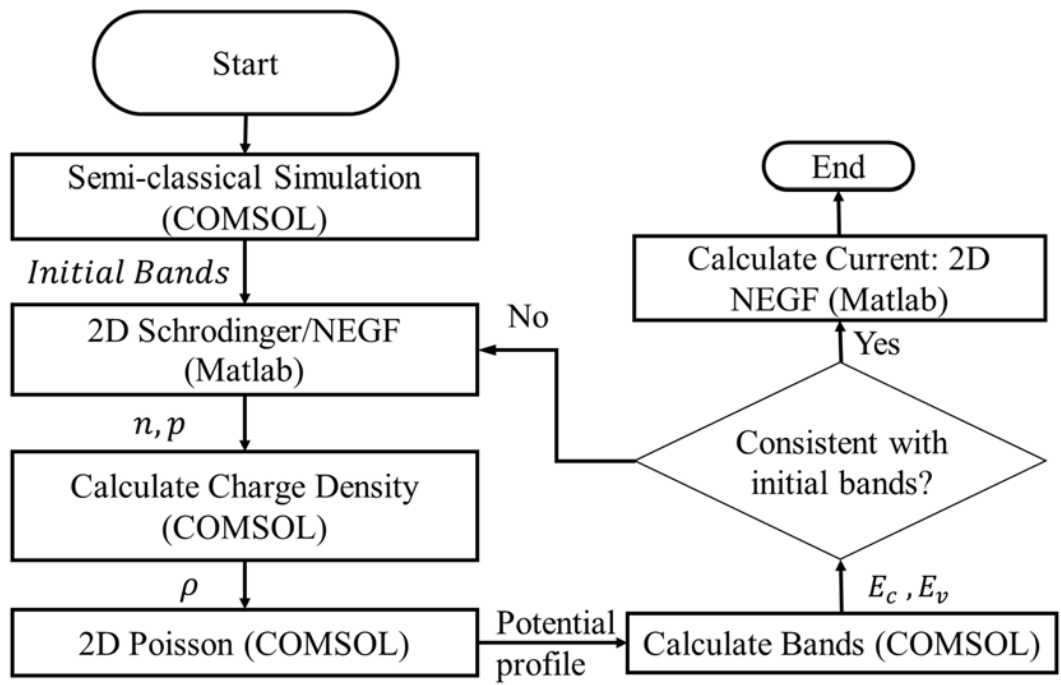

**Figure 4.5: Flowchart for implementation of the simulation process**

The simulation process goes through the following stages for a given bias point i.e for a given value of gate and drain voltages:

Step 1: First a semi-classical simulation based on Poisson's equation and classical drift-diffusion equation is done in COMSOL Multiphysics from which we get the initial prediction of the band profile i.e. $E_c$ and $E_v$.

Step 2: The band profile is imported into Matlab where we calculate the carrier concentrations i.e. $n$ and $p$ using 2D NEGF or closed boundary Schrodinger's equation depending on the required accuracy from the simulation.

Step 3: Then the carrier concentration is imported into COMSOL. When a study is run, charge density $\rho$ is calculated from the carrier concentration and doping profile.

Step 4: Then the COMSOL PDE solver is used to solve the 2D Poisson equation for proper boundary conditions and a new potential profile $V(x, y)$ is found.

Step 5: The COMSOL study generates new band profiles from the calculated potential profile.

Step 6: The band profile is imported into Matlab. If it is within a given tolerance limit of the initial band profile, then the simulator proceeds to next step. Otherwise it returns to Step 2 for another iteration.

Step 7: After convergence is achieved, all the required mathematical operations required to integrate currents for different sub-bands using 2D NEGF is performed in Matlab, from which we get the final value of current (per unit width) for the given bias point.

## 4.8 Chapter Summary

In this chapter we have discussed all the steps necessary to implement our model in a realistic simulator. We have presented a TFET design for experimentation and demonstrated how we performed simulations for this device. In the next chapter we will discuss the results found by carrying out the simulation procedure and discuss their implications.

# Chapter 5
# Results and Discussions

In this chapter we will discuss the results found by carrying out the simulation process described in the previous chapter. We will compare the results with results found from other models of TFET. We will also explain the reason for agreement of mismatch in results of different models.

## 5.1 Comparison of Transmission Coefficient for 2D potential: NEGF and WKB Approximation

Before proceeding to the simulation results let us first demonstrate how the transmission coefficient found from 2D NEGF formalism matches with the WKB approximation for 2D potential distribution under different conditions. As we mentioned in chapter 2, WKB approximation is used in non-local semi-classical models to calculate transmission coefficient and thus the tunneling current. So, the comparison of transmission coefficient will give us insight into how our model, which is based on 2D NEGF, may agree or disagree on the current calculated with static and dynamic path non-local models.

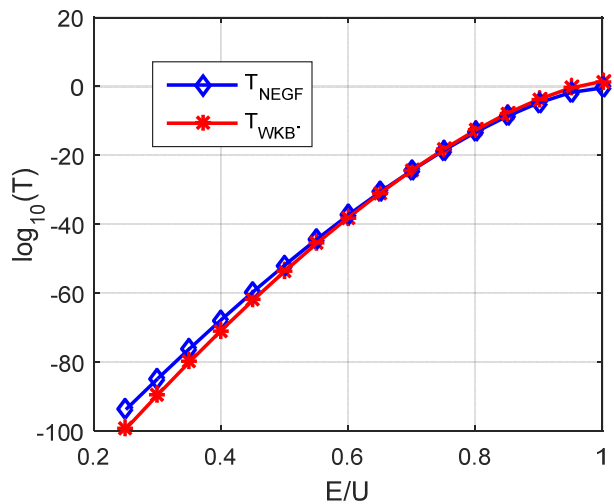

**Figure 5.1: Transmission coefficient for no skewness and no confinement: NEGF vs WKB**

We know that to apply WKB approximation, we need to assume a tunneling direction. For the sake of this comparison we always took the tunneling direction to be in the $x$ direction. The $y$ direction is taken to be the perpendicular direction. The *z* direction is excluded from discussion because the potential is 2D and there is no variation of potential in this direction.

If the potential profile is linear, there is no variation in potential in the *y* direction and tunneling only occurs in the *x* direction as assumed by the WKB approximation. Also if the device dimension is wide along the *y* axis there is no lateral confinement. Comparing the transmission coefficients in this situation, as shown is Figure 5.1Figure 5.2 we see that the 2D NEGF and WKB approximation give almost the same value of transmission coefficient for all energy levels.

Now if we reduce the device dimension along the *y* axis to introduce a high lateral confinement and the compute the transmission coefficient by the two abovementioned approaches, we see from the results shown in Figure 5.2 that NEGF formalism gives less transmission than WKB approximation. This is because, The WKB approximation cannot inherently take into account the transmission of wave function through a confined region without proper modification. So, if a non-local model needs to account for confinement it should be properly modified (for example, as in [40]) before applying WKB approximation.

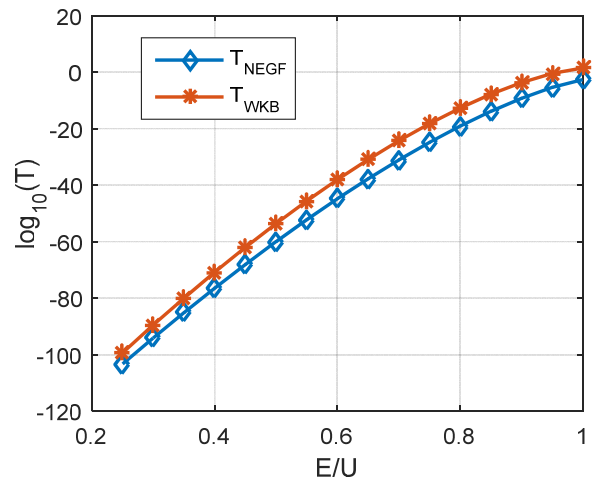

**Figure 5.2: Transmission coefficient for high confinement: NEGF vs WKB**

Now let us modify the potential profile to add variation in the *y* direction. This means that the potential distribution would no longer be linear. For this skewed potential distribution tunneling process does not strictly follow any single preferred direction. If the assumed direction is wrong it would probably yield a tunneling path that is longer than the actual path and thus a lower transmission coefficient than the actual value. This is demonstrated in Figure 5.3 where we see that the WKB approximation gives lower transmission for a skewed potential than 2D NEGF formalism because of wrong assumption of tunneling path. The 2D NEGF formalism on the other hand does not make any of the previous assumptions on tunneling path and thus gives the actual transmission coefficient.

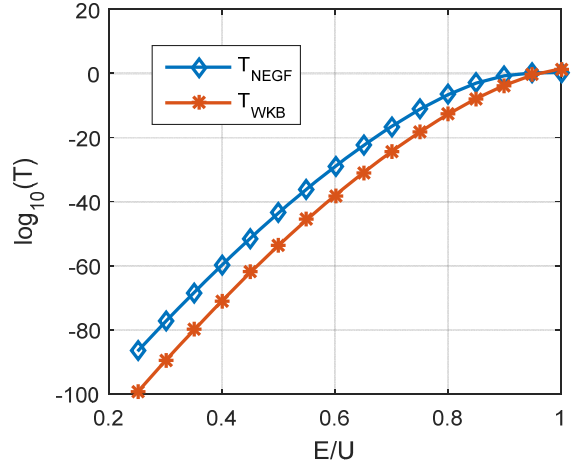

**Figure 5.3: Transmission coefficient for high skewness: NEGF vs WKB**

So, we can conclude from the above discussion that the WKB approximation gives the correct Transmission coefficient only when there is no large variation in potential in the y direction (perpendicular direction) and tunneling occurs only in the x direction (Assumed direction) and there is no confinement. The result can be summarized in Table 5.1.

**Table 5.1: Transmission coefficient found from WKB approximation**

| Condition | Transmission Coefficient from WKB approx. |
|---|:---:|
| **No Confinement and No Skewness** | Actual |
| **High Confinement** | More |
| **High Skewness** | Less |
| **High Confinement and Skewness** | Irregular |

## 5.2 Typical Carrier Distribution for Lateral Confinement Found from 2D NEGF Formalism

Now we want to demonstrate the effect of lateral confinement on charge distribution and how the 2D NEGF formalism predicts this effect. Figure 5.4 and Figure 5.5 shows the electron concentration near the channel region for a device with a body thickness of 5 nm and an n type (donor atom) doping concentration of $2.5 \times 10^{25} \text{cm}^{-3}$ in the drain. We see that, due to lateral confinement the carrier concentration fall off at the edges of the device where closed boundary condition is applied, but not at the lead where open boundary condition is applied.

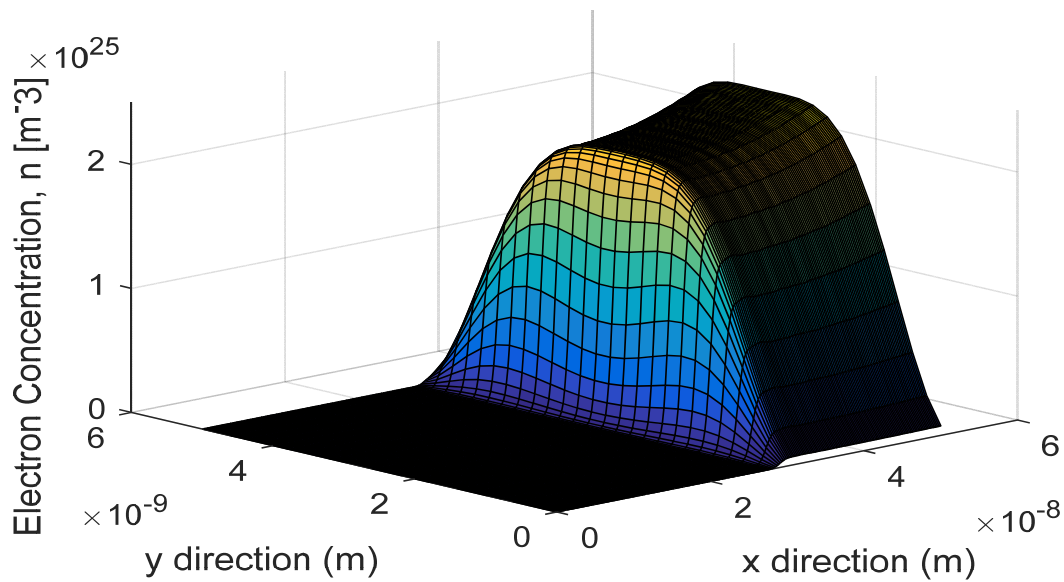

**Figure 5.4: Isometric view: electron concentration near the channel region for a device with a body thickness of 5 nm and a drain (n type) doping concentration of $2.5 \times 10^{25} \text{cm}^{-3}$**

Also, notice that carrier concentration in the barrier falls off rapidly. Excluding the abovementioned regions, the currier concentration is the same as the bulk carrier concentration calculated by semi-classical theory. Also this proves that, band to band tunneling does not change the carrier concentration significantly, because of high barrier in the bandgap.

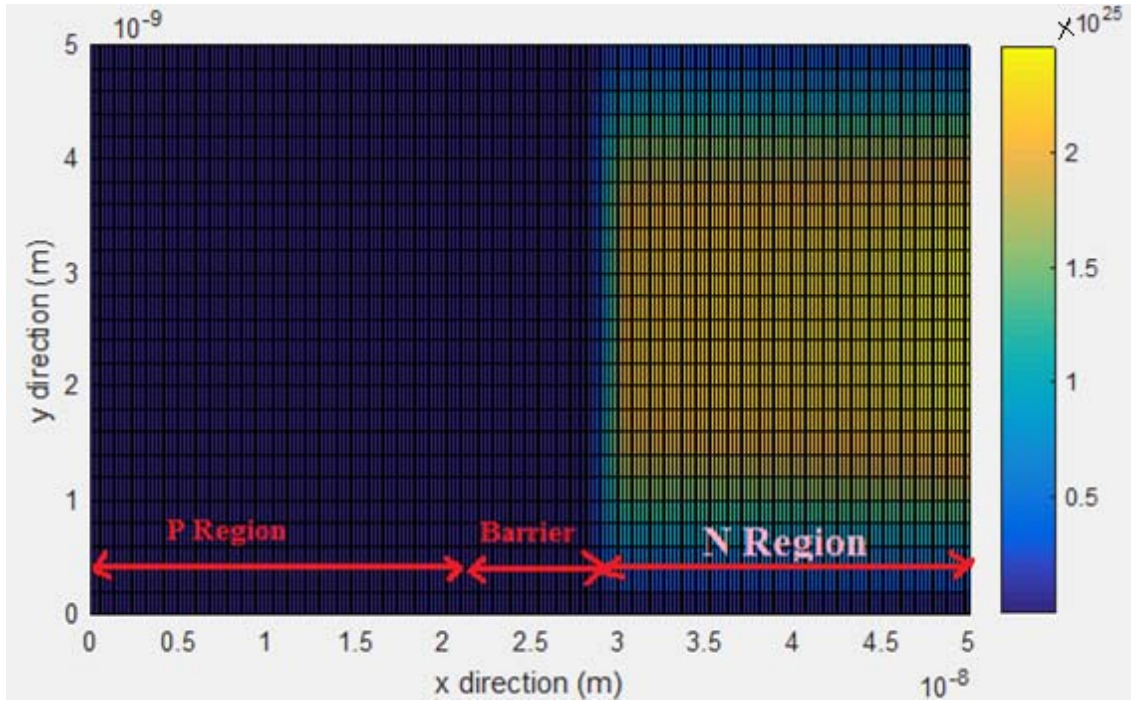

**Figure 5.5: Top view: electron concentration near the channel region for a device with a body thickness of 5 nm and a drain (n type) doping concentration of $2.5 \times 10^{25} \text{cm}^{-3}$**

## 5.3 Simulation Results and Relevant Discussions

Let us now demonstrate the simulation results for the device introduced in section 4.6. Figure 5.6 shows the simulation results found from different models for a drain voltage of 1.5V and body thickness of 10nm. The reason for our choice of these parameters will be explained shortly. As we can see that Hurkx model [55] which is Local model gives a very crude estimation of current. This is to be expected because local models make various simplifying assumptions which make them unsuitable for precise simulations. So, we will exclude local models from the results of the next simulations.

Results for the non-local models are taken from [40] which were done in Silvaco Atlas [56]. Since the body thickness is 10 nm, there is moderate confinement along the lateral direction. As we can see the conventional non-local model cannot account for the lateral confinement and predicts higher current than actual value. But when the non-local model is modified to account for lateral confinement according to [40], it agrees with the simulation result of our model. This is because, our model directly solves the Schrodinger's equation and inherently takes the effect of confinement into account.

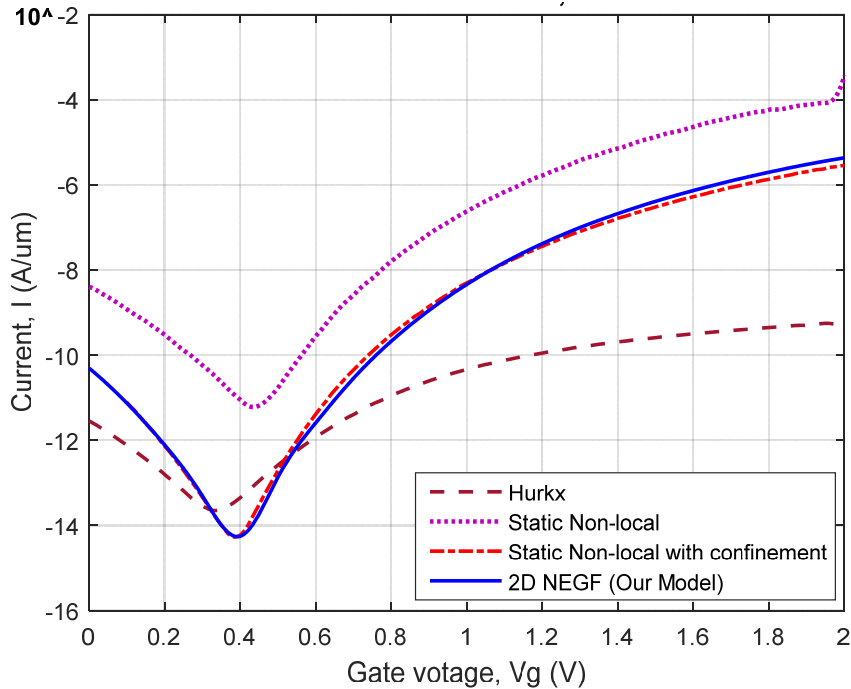

**Figure 5.6: Simulation results found from different models (including our model) for a drain voltage of 1.5V and body thickness of 10nm**

To further investigate the reason for agreement, let us take a look at the band diagrams (Figure 5.7) and the tunneling effective barrier (Figure 5.8 and Figure 5.9). We see that the band profile and consequently the tunneling effective barrier is fairly linear i.e. invariant along the $y$ direction (perpendicular direction). The non-linear model, which uses the WKB approximation to calculate transmission coefficient, assumes a tunneling direction along the x axis.

As we demonstrated in section 5.1, NEGF and WKB approximation predict the same transmission when the barrier is linear, because there is only one possible tunneling direction. As for the effect of confinement, the modification given in [40] takes care of the low amount of confinement present due to the moderately thin body. As a result non-local model is able to predict the actual current and thus agrees with our model.

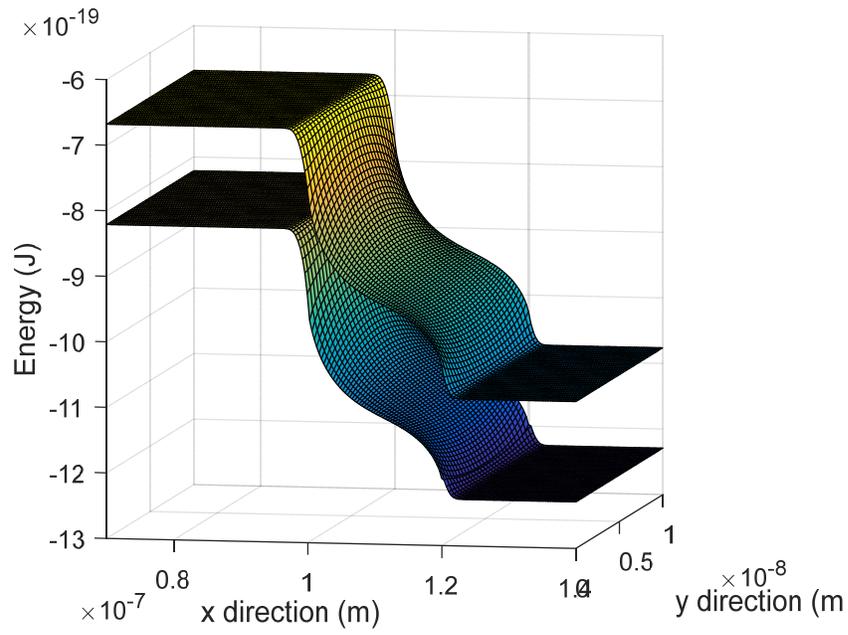

**Figure 5.7: Band profile for a drain voltage of 1.5V and body thickness of 10nm**

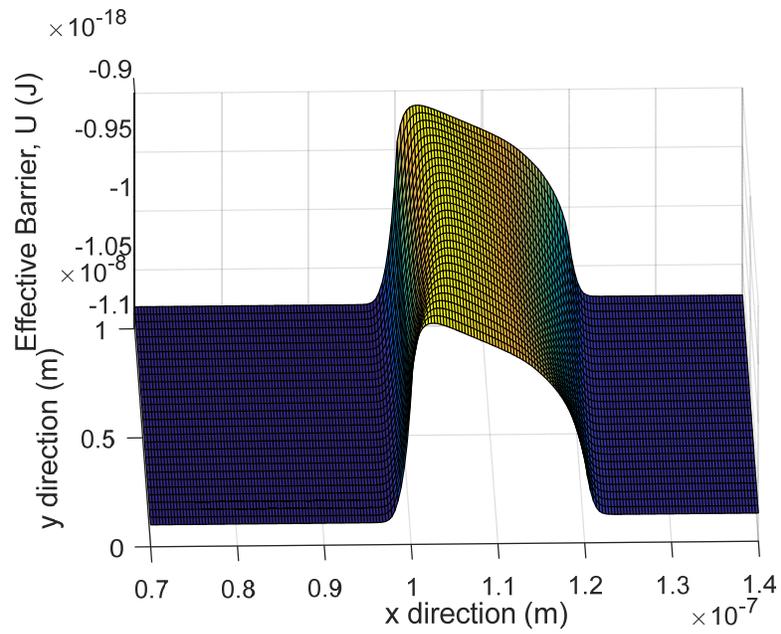

**Figure 5.8: Isometric view: typical tunneling barrier for a drain voltage of 1.5V and body thickness of 10nm**

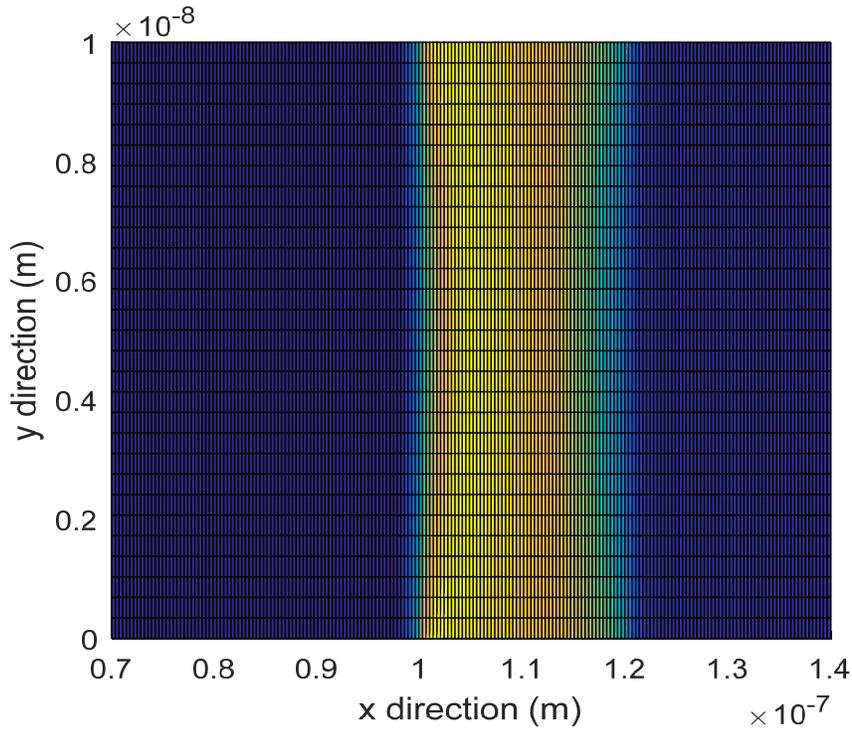

**Figure 5.9: Top view: typical tunneling barrier for a drain voltage of 1.5V and body thickness of 10nm**

Figure 5.10-12 shows the simulation results for a drain voltage of 0.1V and channel thicknesses of 10nm, 5nm and 3nm. We varied the body thickness to see the effect of lateral confinement on current-voltage characteristic. As can be seen from Figure 5.12, our model disagrees with the modified non-local model at higher confinement, that is for lower body thickness. Also our model predicts much more current at higher voltage than the modified non-local model, especially at 10 nm. The modified non-local model shows a downward trend for increasing gate voltage which should not be seen practically. In general our model predicts more current than the model that considers confinement but less current than the model that does not consider confinement.

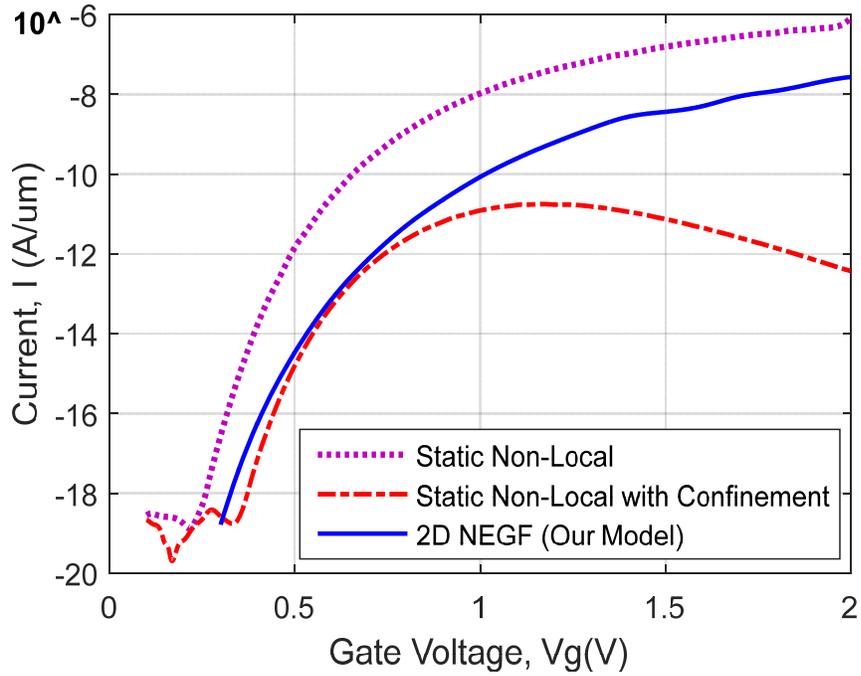

**Figure 5.10: Simulation result for a drain voltage of 0.1V and channel thicknesses of 10nm**

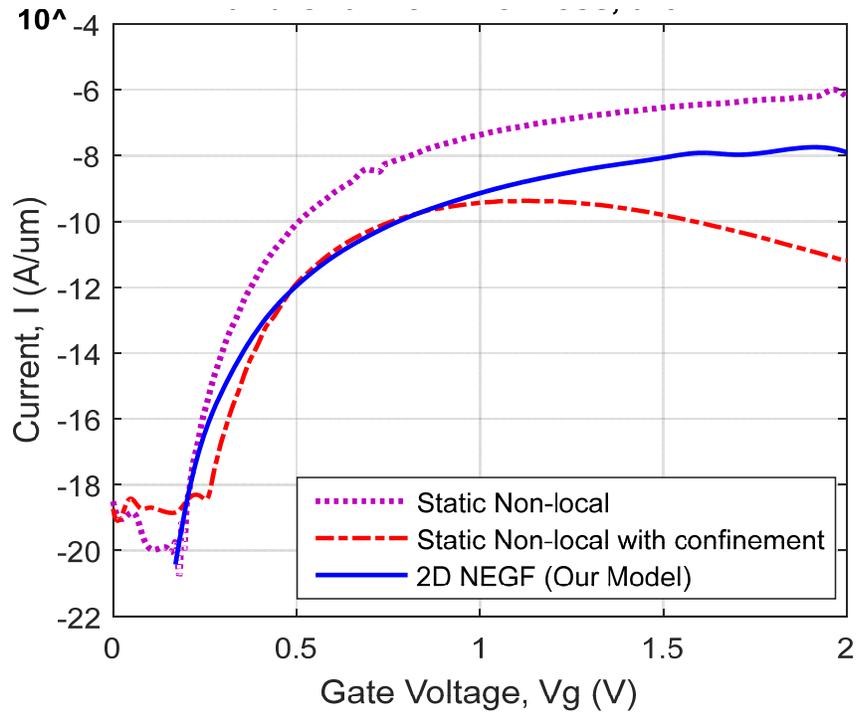

**Figure 5.11: Simulation result for a drain voltage of 0.1V and channel thicknesses of 5nm**

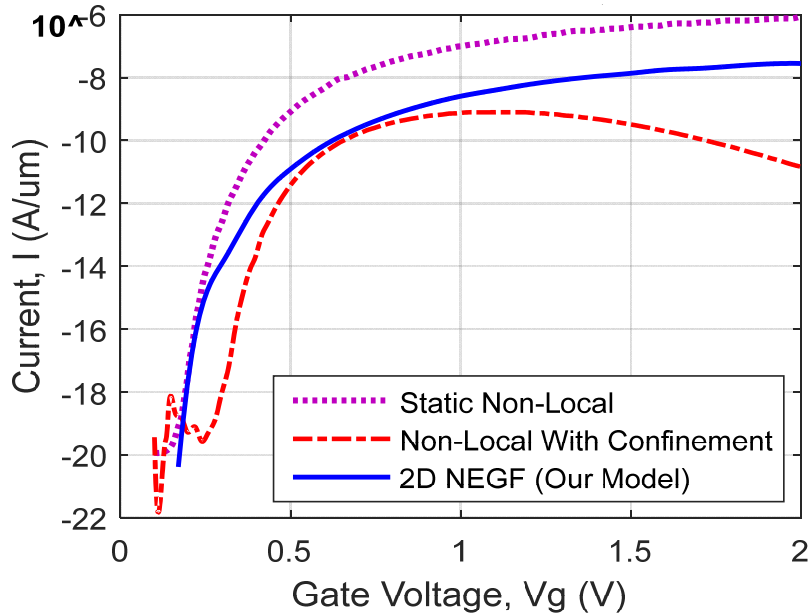

**Figure 5.12: Simulation result for a drain voltage of 0.1V and channel thicknesses of 3nm**

From the discussion in section 5.1 we conclude that the reason for this disagreement can be lateral confinement or skewed potential or a combination of both. First, let us discuss how the modified non-local model handles confinement. As discussed in [40], this model tries to include the effect of confinement by solving the Schrodinger's equation in the $y$ direction to find the energy of the first bound state in both conduction and valence bands. Since there is no transmitting states below this energy level, in this model it is assumed that tunneling takes place between the first bound states in both bands. Which means that confinement is approximated by an increase in effective bandgap in the confined region.

However, this is a highly simplified picture of the actual effect of confinement. Figure 5.13 shows the actual staircase shaped density of states curve for a laterally confined region [57], and how the modified non-local model tries to approximate this curve. While, for lower level of confinement, increased effective bandgap approximation may work, but for higher level of confinement it ignores the abrupt rise in transmitting states at the bottom of the DOS curve, and consequently underestimates the number of transmitting states and ultimately the amount of drain current. So this explains why the modified non-local model gives lower current than our model for a body thickness of 3 nm. However, it does not explain why this model gives much lower current at higher voltage, than our model, which will be investigated next.

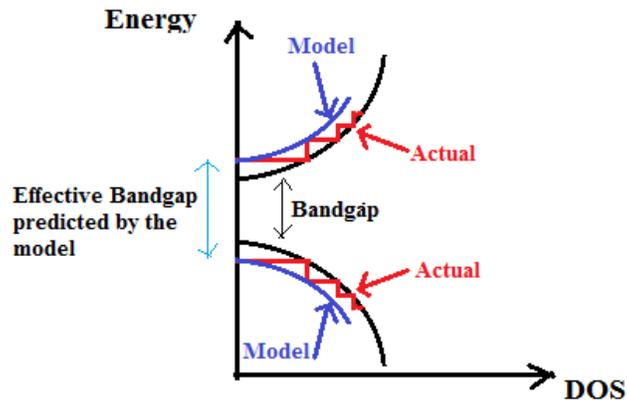

**Figure 5.13: Density of states for unconfined (black) and confined (red) region and DOS approximated by model (blue) given in [40] for confined region**

Figure 5.14 shows the band diagram for simulation done at drain voltage of .1V, gate voltage of 1.1V and a body thickness of 10nm. As seen from the figure, although the device dimension imposes only moderate confinement on the current, a very narrow confined region in the conduction band forms just beneath the gate, just like the inversion layer in MOSFETs. Most of the current flows through this highly confined region. As we mentioned above, the simplifying assumption of the modified non-local model leads it to predict much lower than the actual current. This explains the downward trend in current shown by the modified non-local model for higher gate voltages.

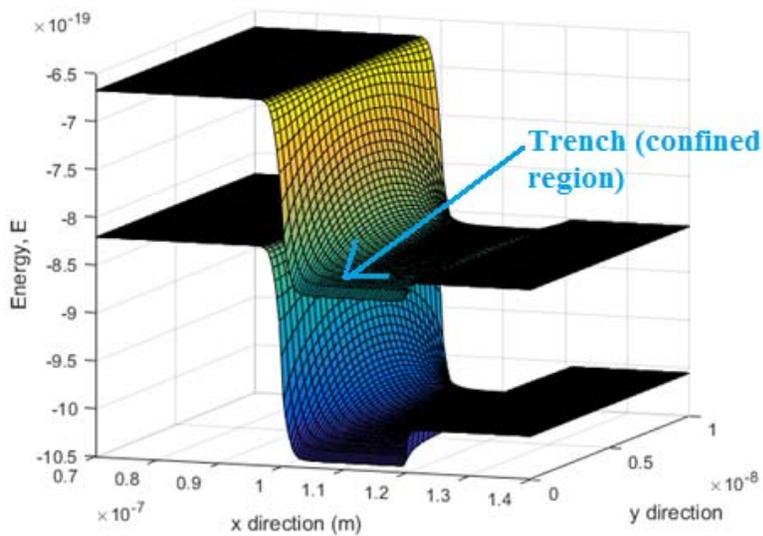

**Figure 5.14: Band diagram for drain voltage= .1V, gate voltage= 1.1V and thickness=10nm**

As we see from Figure 5.10, the difference in current is particularly high at higher gate voltages for body thickness of 10 nm. This can be explained from the skewness of the barrier which increases with body thickness because of the higher variation of the potential along the $y$ direction. Figure 5.15 shows the barrier shape for a gate voltage of 1.1V. We see that the barrier is highly skewed.

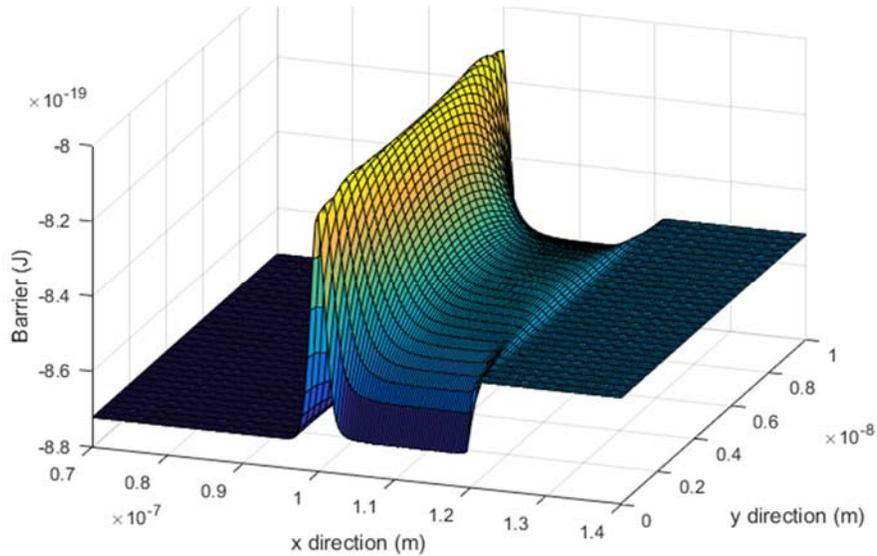

**Figure 5.15: Typical tunneling barrier for drain voltage = .1V, gate voltage = 1.1V and thickness =10nm**

The top view of the barrier shown in Figure 5.16 reveals a wedge shape rather than a linear shape. The non-local model invariantly takes the $x$ direction as the tunneling direction (Figure 5.16(a)). Whereas a set of more probable tunneling direction is shown in Figure 5.16(b) which vary along thickness of the device. As, the non-local model considers the wrong tunneling paths, especially just below the gate where tunneling is highest, it ends up calculating longer tunneling distances, which leads to less transmission and less current. In summary, the non-local model predicts even less current for higher body thickness at higher gate voltages because of its failure to account for the skewness of the potential distribution. Our model does not make any such assumption and predicts the actual transmission and the actual current which is of course more than that predicted by the modified non-local model. However the current is less than the current estimated by the non-local model that does not consider confinement because our model does consider confinement inherently.

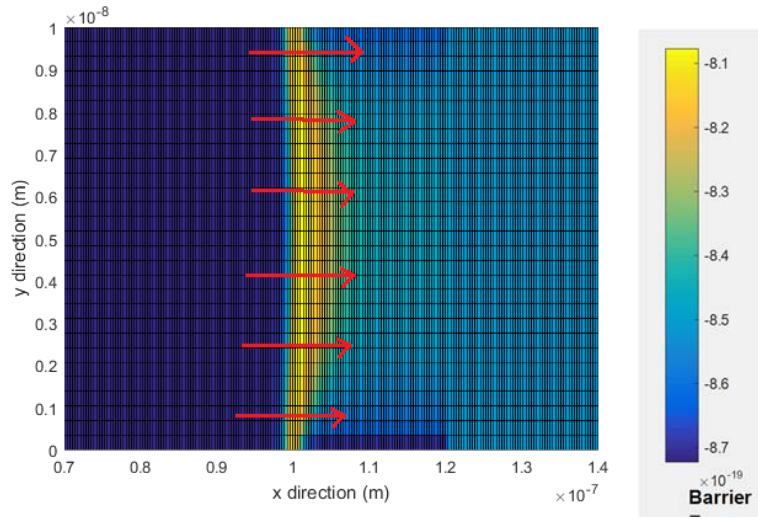

(a)

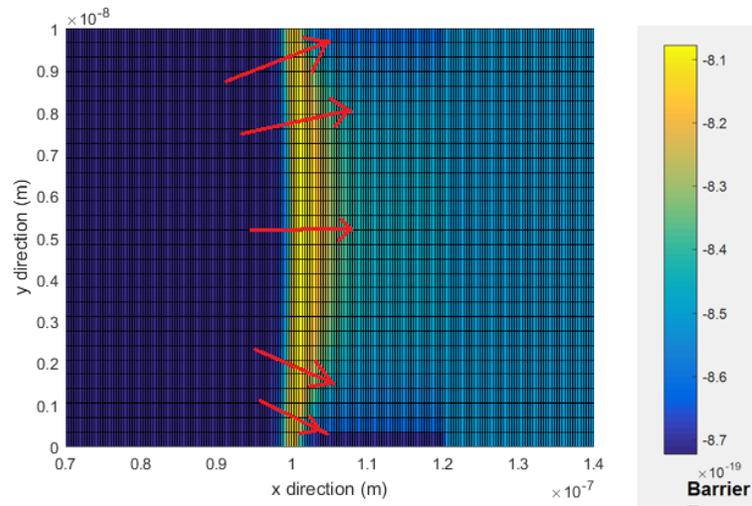

(b)

**Figure 5.16: The top view of the barrier in Figure 5.15 and (a) Assumed tunneling path by the non-local model, (b) Actual tunneling path**

In this section we have discussed how simplifying assumptions may lead to wrong predictions. Our model, on the other hand does not make any such assumption and more suitable for precise simulations.

## 5.4 Benchmarking

Dynamic non-local path models try to account for the skewness of the potential profile by taking the gradient of the bands as tunneling direction. From the above discussions we may conclude that, if there is no confinement and the non-local model can account for the skewness, its predictions should match with that of our model. Our model was used in [58] to simulate a hetero-gate double-gate TFET. The simulation result of our model was benchmarked against that of a dynamic non-local model which was taken as a reference. The device structure is shown in Figure 5.17(a). There was no pronounced lateral confinement in this device, so the simulation result matched well against that of the dynamic non-local model (Figure 5.17(b)).

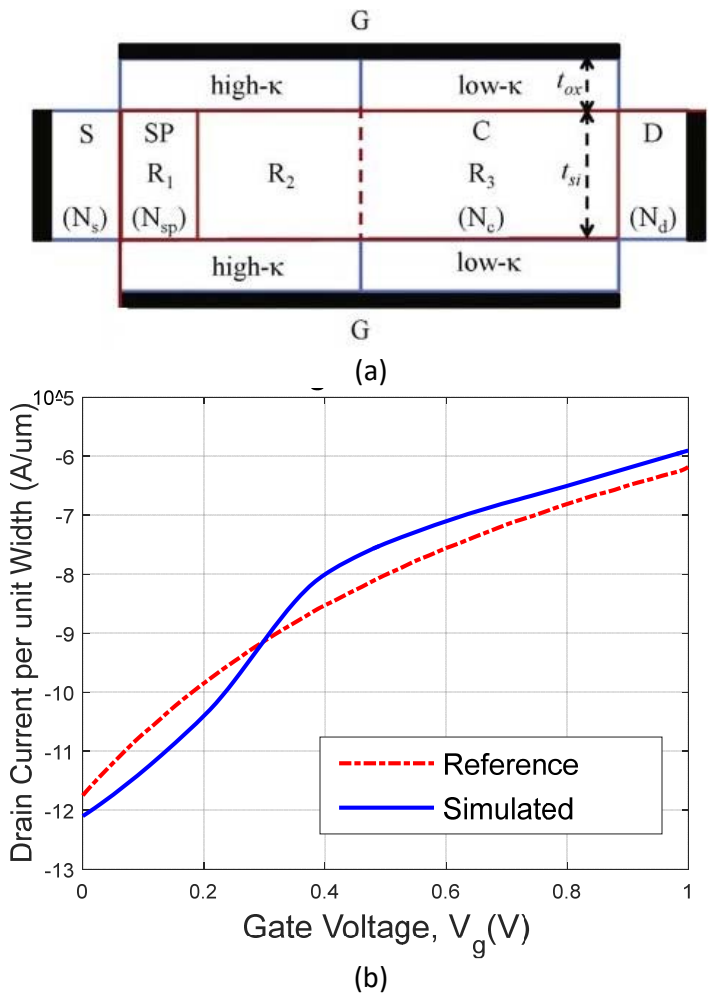

Figure 5.17: Benchmarking: (a) Simulated device (b) Simulation results

## 5.5 Chapter Summary

In this chapter we have discussed the implications of our model i.e. the effects our model can take into account. We have clarified how our model may agree with non-local semi-classical models under certain conditions, and how it overcomes the limitations of semi-classical models.

# Chapter 6
# Conclusions

In this chapter we review and summarize the work done so far. Then we will clarify the limitations of the developed model and point out future prospects of improvement and expansion.

## 6.1 Summary

In this work, we have developed a simplistic quantum mechanical model for BTBT in TFET that is applicable for 2D potential distribution using effective mass approximation and 2-band model. Then, we have integrated the developed model with 2D NEGF formalism for proper treatment of 2D potential distribution. We have simplified the results obtained from NEGF formalism by making some approximations for faster simulation. We have demonstrated the simulation methodology and discussed the pitfalls in simulation and how to tackle them.

Then we implemented our model to simulate a TFET structure. From the results of the simulation we have shown the agreement of our model with existing non-local models at proper conditions. We have also pointed out the limitations of the semi-classical models and shown how our model overcomes the limitations of semi-classical models.

## 6.2 Limitations of Our Model

We should point out some limitations of our model to clarify where our model is applicable.

- Our model uses effective mass approximation. So it cannot be applied where effective mass approximation is not valid, for example in graphene mono-layer or in ultra-thin body devices where the band structure change so much that the effective mass approximation is no longer valid.
- Our model uses an approximate dispersion relation in the bandgap. So it is not as accurate as multiband models, which uses a much more realistic dispersion relation in the bandgap.
- Our model can only be applied when the potential is invariant along one of the axes (in our case the z axis). For 3D potential distribution we would need a 3D model, and the simulation would be much more time consuming.
- Our model gives more accurate result for direct band to band tunneling, because it is based on the assumption of conservation of energy and momentum as discussed in section 3.2.

Consequently, for phonon and trap assisted tunneling it would give less accurate result unless proper elaboration of the model is done.

## 6.3 Prospects of Future Work

We propose some prospects of improvement and elaboration of the model

- In developing the model we have assumed only one valence band for holes. But in most cases there are two degenerate bands with different effective masses. Our model can be extended to include these two bands for heavy holes and light holes.
- For simplicity we have assumed isotropic effective masses in the bands. But most of the crystals have at least 2 different effective masses for carriers in the transport and longitudinal directions. So our model can be modified to include these two different effective masses in different directions.
- We have excluded all types of scattering mechanisms for simplicity and for faster simulation. But they can be easily modeled with Buttiker probes. This also allows us to approximately model thermal generation and recombination.
- Our model basically assumes direct band to band tunneling. The model could be extended to include phonon and trap assisted tunneling.
- Our model was developed keeping specialized quantum mechanical structures like the superlattice structure or modulation doped devices in mind. But we have not verified the applicability of our model for these structures. We intend to do so in a future work.

# Appendix A
# Formulation of the 2D Self Energy Matrix

The formulation of the self-energy matrix is given in [52]. The self-energy matrix is the solution to the Matrix equation

$$-t_x^2 g^2 + (ES - H_{1D})g - S = 0 \quad (A.1)$$

Where S is the identity matrix. This equation can be solved if we diagonalize $(ES - H_{1D})$. Let

$$(ES - H_{1D}) = X = QX'Q^\dagger \quad (A.2)$$

Where, $X'$ is a diagonal matrix. Since $H_{1D}$ is Hermitian, $Q^{-1} = Q^\dagger$.

The equation takes the form

$$-t_x^2 g'^2 + Q^\dagger(ES - H_{1D})Qg - S = 0 \quad (A.3)$$

$$\text{Where, } g' = Q^\dagger g Q \quad (A.4)$$

Now the solution is given by

$$g = Qg'Q^\dagger = \frac{X + Q\sqrt{X'^2 - 4t_x^2 S}Q^\dagger}{2t_x^2} \quad (A.5)$$

$g$ is used to form the self-energy matrix $\Sigma$.

# Appendix B
# Partial Calculation of The Green's Function Matrix

A method of optimizing simulation of 2D NEGF is suggested in [51]. From the formulation of Appendix A, the self-energy matrix can be written as

$$\Sigma = Q\Lambda Q^\dagger \qquad (B.1)$$

$\Lambda$ is formed from the matrix $g'$ in equation (A.5).

Now, the broadening function can be written as

$$\Gamma = Q\Xi Q^\dagger \qquad (B.2)$$

Where,

$$\Xi = j(\Lambda - \Lambda^\dagger) \qquad (B.3)$$

In our case $\Xi$ is a positive real diagonal matrix and $Q$ is a real matrix. So, we can write,

$$\Gamma = Q\sqrt{\Xi}\sqrt{\Xi}Q^T = \Upsilon\Upsilon^\dagger \qquad (B.4)$$

Now we can use this decomposition to simplify the spectral density function which is given by

$$A = G\Upsilon\Upsilon^\dagger G^\dagger = YY^\dagger \qquad (B.5)$$

Where, $Y = G\Upsilon$, which means that

$$\left[ES - H(E) - \sum_\alpha \Sigma_\alpha(E)\right] Y = \Upsilon \qquad (B.6)$$

The column number of $\Upsilon$ is the truncation order in $y$ direction. For high lateral confinement, only a few lowest energy sub-bands are important, so only a few columns of $Y$ are needed to obtain $A$ by solving equation (B.6) instead of calculating the whole Green's function matrix $G$.

Similarly, optimization for transmission coefficient can be done using the decomposition given in equation (B.4). After making these optimizations, the simulation becomes much faster, especially for lateral confinement, i.e. thin body devices.